\documentclass[10pt,pre,twocolumn,showpacs,floatfix,aps]{revtex4-1}     
\usepackage{graphicx}           

\usepackage{amssymb}
\usepackage{amsmath}
\usepackage{natbib}
\usepackage{color}
\usepackage{eucal}
\newcommand{\mthr}{\mathrm}

\begin{document}
\title{Translational and rotational dynamics of colloidal particles interacting through reacting linkers }
\author{Pritam Kumar Jana}
\email{Pritam.Kumar.Jana@ulb.ac.be} 
\affiliation{Center for Nonlinear Phenomena and Complex Systems, Code Postal 231, Universite Libre de Bruxelles, Boulevard du Triomphe, 1050 Brussels, Belgium}
\author{Bortolo Matteo Mognetti}
\email{bmognett@ulb.ac.be} 
\affiliation{Center for Nonlinear Phenomena and Complex Systems, Code Postal 231, Universite Libre de Bruxelles, Boulevard du Triomphe, 1050 Brussels, Belgium}

\begin{abstract}
Much work has studied effective interactions between micron--sized particles carrying linkers forming reversible, inter-particle linkages. These studies allowed understanding the equilibrium properties of colloids interacting through ligand--receptor interactions. Nevertheless, understanding the kinetics of multivalent interactions remains an open problem. Here, we study how molecular details of the linkers, such as the reaction rates at which inter--particle linkages form/break, affect the relative dynamics of pairs of cross-linked colloids. Using a simulation method tracking single binding/unbinding events between complementary linkers, we rationalize recent experiments and prove that particles' interfaces can move across each other while being cross-linked. We clarify how, starting from diffusing colloids, the dynamics become arrested when increasing the number of inter-particle linkages or decreasing the reaction rates. Before getting arrested, particles diffuse through rolling motion. The ability to detect rolling motion will be useful to shed new light on host-pathogen interactions.
\end{abstract}

\maketitle

\textbf{Introduction.} 
Understanding the dynamics of colloids interacting through ligand/receptor (or linker) molecules forming inter-particle linkages is a pivotal challenge in nanoscience. For instance, particles functionalized by reactive linkers are currently used to self-assemble crystals \cite{park2008dna,auyeung2014dna,rogers-manoharan,Ducrot_NatMat_2017,Wang_NatComm_2017,Liu_Science_2016} or disordered structures \cite{Biffi_PNAS_2013,di2014aggregation,Jasna_PRL_2018} featuring sought properties and responsive behaviors. Secondly, ligand-receptor interactions are broadly employed by biology to regulate cell adhesion and inter--membrane trafficking \cite{Alberts}. The level of complexity and functionality achieved in these biological systems is out of the reach of current in vitro designs, partially because of a still incomplete understanding of multivalent interactions \cite{knorowski2011materials,A-UbertiPCCP2016}. 
\\
Many investigations have studied effective interactions between particles forming reversible linkages \cite{BellBJourn1984,melting-theory1,RogersPNAS2011,MladekPRL2012,francisco-pnas,TitoJCP2016,PhysRevE.96.012408,jenkins2017interaction,DiMichelePRE2018}. 
Effective interactions account for entropic terms (such as the avidity of the system) through ensemble averages over all possible linkages featured by the system at a given colloids' configuration and have been used to study systems with micron sized particles not directly accessible to Molecular Dynamics simulations \cite{li2012modeling}. However, effective interactions cannot quantify the kinetics of multivalent interactions which, often, is pivotal in controlling the outcome of an experiment. For instance, most of the experiments using DNA sticky--ends \cite{mirkin,alivisatos} tethered to micron--sized colloids tempt to self-assemble disordered aggregates \cite{melting-theory2} rather than crystalline structures. Arrested aggregates are the result of sticky interactions, as corroborated by dedicated experiments showing how the relative diffusion between DNA coated surfaces proceeds through hopping steps happening during intervals in which the surfaces are free from any inter--particle linkage \cite{chaikin-subdiffusion,rogers2013kinetics}. Such finding implies that bound colloids diffuse only in a narrow range of temperatures close to the melting temperature. However, more recent experiments employing densely functionalized colloids have reported diffusive dynamics over broader temperature ranges with the possibility for the colloids to move against each other while being cross--linked \cite{KimCrockerLang2006,wang2015crystallization,wang2015synthetic}. 
\\
In this work, we study the relative diffusion of two 1$\,\mu$m diameter colloids functionalized by up to $5\cdot 10^4$ ligands \cite{wang2015crystallization,wang2015synthetic} using a numerical framework tracking the reaction dynamics of each linker. We find that particles' interfaces can move past each other while being bound when the number of inter-particle linkages is sufficiently low or the rates of forming/breaking linkages are sufficiently high. The latter finding rationalizes the results of Ref.~\cite{wang2015crystallization,wang2015synthetic} given that at higher coating densities the rates of breaking linkages at the melting transition are higher. The mobility of particles forming few linkages underlies a recent breakthrough reporting crystals of nanoparticles stabilized by smaller, conjugated nanoparticles roaming through the lattice \cite{Girard1174}. 
We also show how particles feature different internal dynamics depending on the balance between the number of linkages and the value of the rates of reaction. In particular, while decreasing the latter, colloids first start rolling \cite{C8SM01430B}, and then stop moving (arrested dynamics). Our findings will be useful to optimize the design of functionalized particles and will provide a valuable tool to study complex cell-pathogen interactions in biology \cite{Delgusteeaat1273,parveendetachment,vahey2019influenza,cicuta2019}. While the present manuscript studies the dynamics of bound colloids, previous contributions have already clarified how, at low reaction rates, aggregation is limited by the diffusivity of the particles  \cite{PetitzonSoftMatter2016,lanfranco2019kinetics}.

\begin{figure*}
\begin{center}
\includegraphics[scale=0.23]{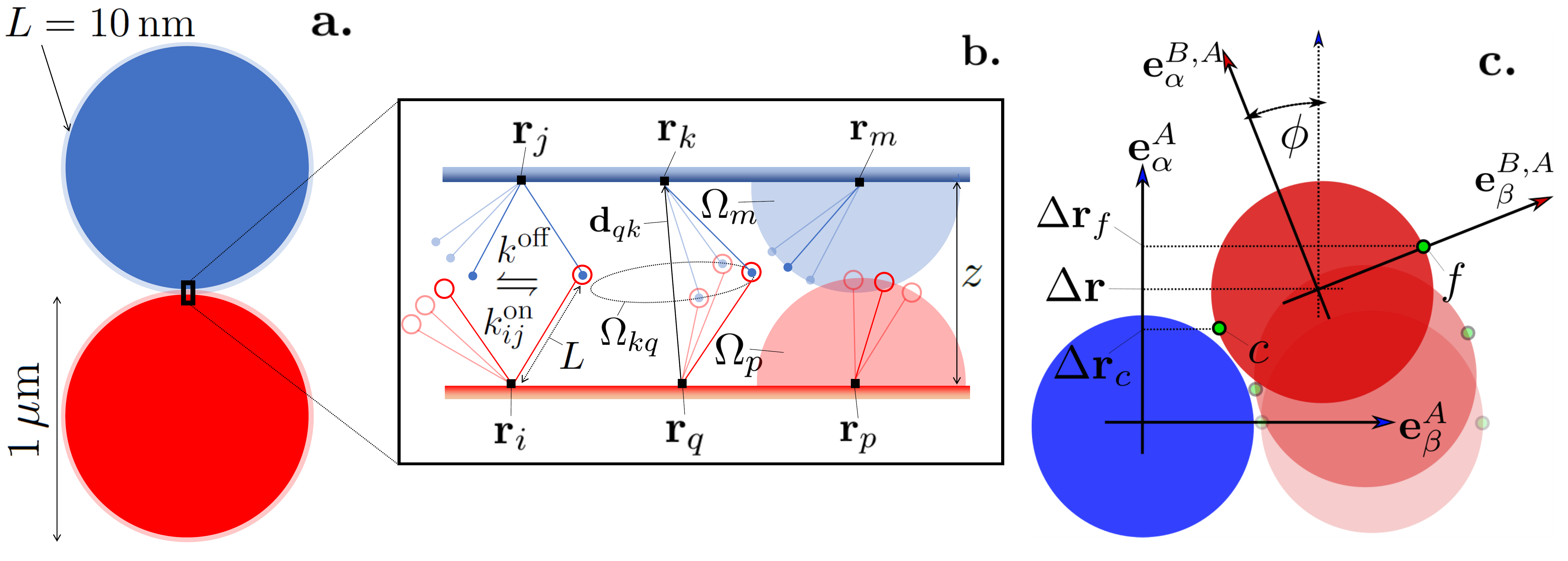}
\caption{ {\bf The system and the modeling strategy.} {\bf a.}~We study micron--sized colloids decorated with $N_b$ linkers of length $L=10\,$nm forming reversible, inter-particle linkages. {\bf b.}~We map linkers onto thin rods tipped by reacting sticky ends (circles). 
$k^\mathrm{on}$ and $k^\mathrm{off}$ are, respectively, the rates at which linkages form and break and are a function of the binding free energy of the sticky ends free in solution ($\Delta G_0$) and the configurational space available to bound and free linkages (respectively, $\Omega_{kq}$ and $\Omega_m$, $\Omega_p$). $d_{qk}$ (similarly $d_{ij}$) is the distance between two bound tethering points ($d_{qk}=|{\bf r}_k -{\bf r}_q|$) and $z$ the local distance between interfaces. {\bf c.} ${\bf e}^A_\alpha$ and ${\bf e}^{B,A}_\alpha$ ($\alpha=1,\, 2,\, 3$) denote the internal reference frame, respectively, of particle $A$ and particle $B$, the latter written using the internal reference of $A$. ${\bf \phi}$ denotes the relative rotation between $B$ and $A$ (see the text and Ref.~\cite{hunter2011tracking} for the definition) while $\Delta {\bf r}$ the lateral displacement of the center of mass of particle $B$ measured in $A$ reference frame. We also track the lateral displacement of the closest ($c$) and farthest point ($f$) of $B$ from $A$ as measured at the starting time.   
}\label{Fig:Fig1}
\end{center}
\end{figure*}
\textbf{The model.}
We consider two colloids with a diameter equal to 1$\,\mu$m \cite{wang2015synthetic,wang2015crystallization,KimCrockerLang2006,RogersPNAS2011} (Fig.~\ref{Fig:Fig1}{\bf a}) each of which is decorated with $N_p$ complementary linkers tipped by reactive sites (full and open circles in Fig.~\ref{Fig:Fig1}{\bf b}).  We map each linker (labeled with $i$, $i=1,\cdots ,N_p$) into a thin, rigid rod of length $L=10\,$nm free to pivot around its tethering point ${\bf r}_i$ (Fig.~\ref{Fig:Fig1}{\bf b}). This representation provides analytic expressions of the force and torque acting on the colloids. Existing literature (e.g.~\cite{melting-theory1,A-UbertiPCCP2016}) has extensively used thin rods to model linkers made of double-stranded DNA terminated by a short single-stranded DNA oligomer (the reactive site, see Fig.~\ref{Fig:Fig1}{\bf b}). The tethering points are uniformly distributed over the spheres' surfaces and do not move. We average our results over more than 50 different coating realizations 
sampled from independent, uniform distributions. Notice that for the largest employed values of $N_p$ ($N_p=5\cdot 10^4$),  the average distance between two neighbouring tethering points (which is equal to $\sqrt{4 \pi R^2/N_p} =7.9\,$nm) is large enough to neglect non-specific (e.g. electrostatic or steric) interactions between linkers.  
\\
The timescales regulating colloidal dynamics at the molecular scale are the rates at which complementary linkers bind and unbind ($k^\mathrm{on}_{ij}$ and $k^\mathrm{off}$ in Fig.~\ref{Fig:Fig1}{\bf b}). Coherently with what done in recent studies modeling DNA-mediated interactions \cite{ParoliniACSNano2016,jana2019surface,PetitzonSoftMatter2016}, we assume that the rates of breaking a linkage is not a function of the position of the tethering points and is equal to the denaturation rate of paired sticky ends when free in solution. The fact that at the thermal scale the forces exerted by the tethering points onto paired sticky ends are not sufficiently strong to deform the structure of the double helix \cite{Ho_BiophJ_2009} justifies the choice of using configuration independent off rates. This result does not apply to other types of ligand-receptor complexes \cite{chang1996influence,chang2000state,shah2011modeling}. Given $k^\mathrm{off}$, we calculate $k^\mathrm{on}_{ij}$ from the hybridization free energy, $\Delta G^\mathrm{hyb}_{ij}$, using $ k^\mathrm{on}_{ij} = k^\mathrm{off} \exp[-\beta \Delta G^\mathrm{hyb}_{ij}]$, where $\beta=1/(k_BT)$, and $T$ and $k_B$ are the temperature and the Boltzmann constant, respectively. $\Delta G^\mathrm{hyb}_{ij}$ comprises the hybridization free energy of free sticky ends in solution, $\Delta G_0$ \cite{santalucia,zadeh2011nupack}, and a configurational part $\Delta G^\mathrm{conf}=k_B T \log ( \Omega_{ij}/ \rho_0 \Omega_i \cdot \Omega_j )$, where $\Omega_{ij}$ and $ \Omega_i \cdot \Omega_j$ are, respectively,  the configurational space available to linker $i$ and $j$ when bound, and when free. 
If $d_{ij}$ and $z_i$ are, respectively,  the distance between the tethering points of two bound linkers and between a tethering point of a free linker and the facing interface (see Fig.~\ref{Fig:Fig1}{\bf b}), we have \cite{MognettiSoftMatt2012,VarillyJCP2012}
\begin{eqnarray}
\Omega_{ij} = {2 \pi L^2 \over d_{ij}} \chi \,  & \qquad & 
\Omega_i = 2 \pi L \cdot \mthr{min}[z_i,L].
\label{Eq:conf1}
\end{eqnarray}
$\Omega_{ij}$ is proportional to the length of the spherical arc traced by the reacted tips of the linkers times a Jacobian term while $\chi$ is the fraction of this spherical arc not excluded by the walls of the colloids. Ref.~\cite{MognettiSoftMatt2012} reports the explicit expression for $\chi$. To obtain portable expressions, we employ a mean field simplification in which we calculate $\chi$ by averaging over the possible lateral distances between tethering points. $\chi$ then reads as follows 
\begin{eqnarray}
\chi (z) = \left\{
\begin{array}{ll}
1 & \mthr{if}\, \, L< z < 2 L
\\
{z \over 2 L - z } & \mthr{if}\, \, z < L
\end{array}
\right.
\label{Eq:conf2}
\end{eqnarray}
where $z=(z_i+z_j)/2$. Notice that after averaging over the lateral positions of the tethering points, the reacted tips are uniformly distributed along the direction orthogonal to the interfaces. $\chi$ is then the fraction of tips-to-plane distances not excluded by the colloids.
The final expression for $\Omega_{ij}$ follows from Eq.~\ref{Eq:conf1} and Eq.~\ref{Eq:conf2}.
The  {\em on} and {\em off} rates read as follows \cite{ParoliniACSNano2016,jana2019surface,PetitzonSoftMatter2016}
\begin{eqnarray}
k^\mthr{off} = \exp[\beta \Delta G_0] \rho_0 k^\mthr{}_0
&\qquad \quad&
k^\mthr{on}_{ij} = \frac{\Omega_{ij}}{\Omega_{i}\Omega_{j}} k^\mthr{}_0 
\label{Equation1}
\end{eqnarray}
where $k_0$ is the binding rate of free sticky ends in solution and we set $\rho_0$ to the standard concentration, $\rho_0=0.62\cdot$nm$^{-3}$.
In this work, we use $k_0$ as a parameter to change the speed of the reactions without affecting equilibrium quantities (e.g., the  number of linkages). Our results are then not specific to DNA linkers but apply to other complexes featuring different $k_0$ \cite{peck2015rapid}. The methodology could also be adapted to include possible cooperativities between binding events as characterized in Molecular Dynamics simulations \cite{randeria2015controls}. 
\\
We define by ${\bf f}^{(A)}_{\langle ij \rangle}$ the effective force exerted on colloid $A$ by the linkage formed by linker $i$ and $j$, labeled with $\langle ij \rangle$ ($ {\bf f}^{(B)}_{\langle ij \rangle}=-{\bf f}^{(A)}_{\langle ij\rangle}$). $ {\bf f}^{(A)}_{\langle ij\rangle}$ accounts for variations of the single linkage partition function, $\Omega_{ij}$, and reads as ${\bf f}^{(A)}_{\langle ij\rangle}=k_B T \nabla_{{\bf r}_A} \log \Omega_{ij}$, where ${\bf r}_A$ is the center of mass of colloid $A$. Using Eqs.~\ref{Eq:conf1}, \ref{Eq:conf2} we find
\begin{eqnarray}
\beta {\bf f}^{(A)}_{\langle ij \rangle} 
&=& 
\left\{
\begin{array}{ll}
-{ \hat {\bf u}_{{\bf d}_{ij}} \over  d_{ij}} & \mthr{if}\, \, L< z < 2 L
\\
-{ \hat {\bf u}_{{\bf d}_{ij}} \over  d_{ij}} +  {\hat {\bf u}}_z \left( {1 \over z}
+ {1 \over 2L -z}\right)  & \mthr{if}\, \, z < L
\end{array}
\right.
\label{Equ:Forceij}
\end{eqnarray}
where $\hat {\bf u}_{{\bf d}_{ij}}$ and ${\hat {\bf u}}_z$ are the unit vectors, respectively, pointing in the direction of ${\bf d}_{ij}$ and orthogonal to the surface of the colloids. For large particles ($R/L\gg 1$), ${\hat {\bf u}}_z$ is approximated by the center-to-center unit vector ${\hat {\bf u}}_z = ({\bf r}_A - {\bf r}_B)/|{\bf r}_A - {\bf r}_B|$. Similarly, each unbound linker $i$ (either tethered to A or B colloid), labeled with $\langle i \rangle$, exerts a force on colloid A equal to ${\bf f}^\mathrm{(A)}_{\langle i \rangle}=k_B T \nabla_{{\bf r}_A} \log \Omega_i$ which reads as follows
\begin{eqnarray}
\beta {\bf f}^{(A)}_{\langle i \rangle}&=&
\left\{
\begin{array}{ll}
0 & \mthr{if}\, \, L< z < 2 L
\\
{ {\hat {\bf u}}_z \over z_i}& \mthr{if}\, \, z < L
\end{array}
\right.
\label{Eq:fi}
\end{eqnarray}
(a similar expression follows for $\beta {\bf f}^{(B)}_{\langle i \rangle}$).
Notice how ${\bf f}^{(A)}_{\langle ij \rangle}$ comprises an attractive contribution along $\hat {\bf u}_{{\bf d}_{ij}}$ and a repulsive term along the centers of mass direction ${\hat {\bf u}}_z$ due to entropic compression. ${\bf f}^{(A)}_{\langle i \rangle}$ only includes a repulsive term along ${\hat {\bf u}}_z$. 
The hard-core interaction between colloids is modeled using the repulsive interaction engendered by a uniform distribution of inert linkers of length $0.75\cdot L$ and reads as follows 
\begin{eqnarray}
\beta {\bf f}^{(A)}_\mathrm{rep} &=& 2\pi \mathrm{R} \sigma \log\left(\frac{0.75L}{D}\right) \mathbf{\hat {\bf u}}_z 
\label{Eq:rep}
\end{eqnarray}
where $\sigma$ and $D$ are, respectively, the density of the inert linkers ($\sigma=0.003$nm$^{-2}$) and the face-to-face distance ($D=|{\bf r}_A-{\bf r}_B|-2\mathrm{R}$). A similar expression follows for ${\bf f}^{(B)}_\mathrm{rep}$. Notice that Eq.~\ref{Eq:rep} follows from a Derjaguin approximation and Eq.~\ref{Eq:fi}. Inert linkers are often used in experiments to stabilize colloidal suspensions.

The total force, ${\bf f}^{(A)}$, and torque, $\boldsymbol{\tau}^{(A)}$, acting on particle $A$ are then equal to $\mathbf{f}^{(A)}=\sum_{<ij>}{\mathbf{f}_{<ij>}^{(A)}}+\sum_{<i>} \mathbf{f}_{<i>}^{(A)}$ and $\boldsymbol{\tau}^{(A)}=\sum_{<ij>} (\mathbf{r}_i-\mathbf{r}_{A}) \times \mathbf{f}_{<ij>}^{(A)}$, where the sums are taken over the list of linkages and free linkers. 
Notice that $\Delta G^\mathrm{hyb}_{ij}$ and ${\bf f}^{(A)}$ are not a function of the reference density $\rho_0$.
We simulate changes in the position, $\Delta {\bf r}_A$, and orientation, $\Delta \boldsymbol{ \varphi}_A$, of the colloids using a Brownian Dynamics scheme \cite{PhysRevE.50.4810} (similar equations hold for $\Delta {\bf r}_B$ and $\Delta \boldsymbol{ \varphi}_B$):
\begin{eqnarray} 
\Delta {\bf r}_A={\mathbf{f}^{(A)} D^T \Delta t\over k_BT} + \delta \mathbf{r} ;
&\,\,\,&
\Delta \boldsymbol{ \varphi}_A={\boldsymbol{\tau}^\mathrm{(A)} D^{R} \Delta t\over k_BT}+\delta \boldsymbol{\varphi}
\label{Equation2}
\end{eqnarray}
We sample $\delta \mathbf{r}$ and $\delta \boldsymbol{\varphi}$ from a Gaussian distribution with zero mean and variance equal to, respectively, $2D^T\Delta t$ and $2D^R\Delta t$, where $\Delta t$ is the integration time step ($\Delta t=2\,$ns). $D^T$ and $D^R$ are the translational and rotational diffusion coefficients given by the Stokes--Einstein--Debye relations: 
\begin{eqnarray}
 D^\mathrm{T}=\frac{k_BT}{6\pi\eta \mathrm{R}}, 
 \qquad
D^\mathrm{R}=\frac{k_BT}{8\pi\eta \mathrm{R}^3},
 \nonumber
\end{eqnarray}
where $k_B$ is the Boltzmann constant, $T$ the temperature ($T=298$\,K), and $\eta$ the dynamic viscosity ($\eta=0.91\,$ mPa$\cdot$ s). These figures lead to $D^T=4.7\cdot10^{5}$nm$^2$s$^{-1}$ and $D^R=1.4\,$  rad$^2$s$^{-1}$.

At each step of the simulation dynamics, we first update the list of linkages $\langle ij \rangle$ using the Gillespie algorithm \cite{gillespie1977exact} as done in Refs.~\cite{jana2019surface,PetitzonSoftMatter2016}. 
At a given colloid configuration, $\{ {\bf r}_A, \, {\bf e}^A_\alpha,\, {\bf r}_B, \, {\bf e}^B_\alpha \}$, we start calculating all the {\em on}/{\em off} rates  of making/breaking linkages between all possible linkers, $k^\mathrm{on}_{ij}$ and $k^\mathrm{off}_{ij}$,  as derived in Eq.~\ref{Equation1}. We then calculate the affinity of a reaction to happens as
\begin{eqnarray}
a_\mathrm{on}^{ij}= \delta_{\langle i\rangle} \delta_{\langle j\rangle} k_\mathrm{on}^{ij}, 
 \quad
 a_\mathrm{off}^{ij}=\delta_{\langle ij\rangle} k_\mathrm{off}^{ij},
\label{Eq:affinities}
\end{eqnarray}
where $\delta_{\langle i\rangle}=1$ ($\delta_{\langle j\rangle}=1$) if linker $i$ ($j$) is free ($\delta_{\langle i\rangle}=\delta_{\langle j\rangle}=0$ otherwise), and $\delta_{\langle ij\rangle}=1$ if a linkage between $i$ and $j$ is present ($\delta_{\langle ij\rangle}=0$ otherwise). Notice that we treat each linker as an independent chemical species.  
We then fire one within all possible reactions with probability 
\begin{eqnarray}
p_\mathrm{on}^{ij}= {a_\mathrm{on}^{ij}\over a_\mthr{tot} }, 
 \quad
 p_\mathrm{off}^{ij}={ a_\mathrm{off}^{ij} \over a_\mthr{tot} },
\end{eqnarray}
where $ a_\mthr{tot}$ is the sum of all affinities. Along with choosing one possible reaction, we sample the time for it to happen ($\tau$), distributed as $P(\tau) =a_\mthr{tot} \exp[-a_\mthr{tot}\tau]$, and increment a reaction clock $\tau_\mthr{reac}$ by $\tau$. If $\tau_\mthr{reac}< \Delta t$ (where $\Delta t$ is the simulation step), we update $\delta_{\langle i\rangle}$ and $\delta_{\langle ij\rangle}$, recalculate the affinities (Eq.~\ref{Eq:affinities}), and fire a new reaction until reaching $\Delta t$. At that point we calculate forces using Eqs.~\ref{Equ:Forceij}, \ref{Eq:fi}, \ref{Eq:rep} and update $\{ {\bf r}_A, \, {\bf e}^A_\alpha,\, {\bf r}_B, \, {\bf e}^B_\alpha \}$ using  Eq.~\ref{Equation2}. The choice of firing more than a single reaction between consecutive configurational updates was motivated by the need of having an affordable algorithm at high reaction rates. Understanding potential bias arising from this limit deserves future dedicated investigations. As compared to existing methods \cite{chang1996influence,chang2000state,shah2011modeling,schoneberg2013readdy}, our algorithm allows simulating configurations in which linkers can potentially bind multiple partners. We neglect hydrodynamic interactions consistently with the fact that the latter are not considered when modeling the linker dynamics. 

\begin{figure*}
\centering
  \includegraphics[scale=0.35]{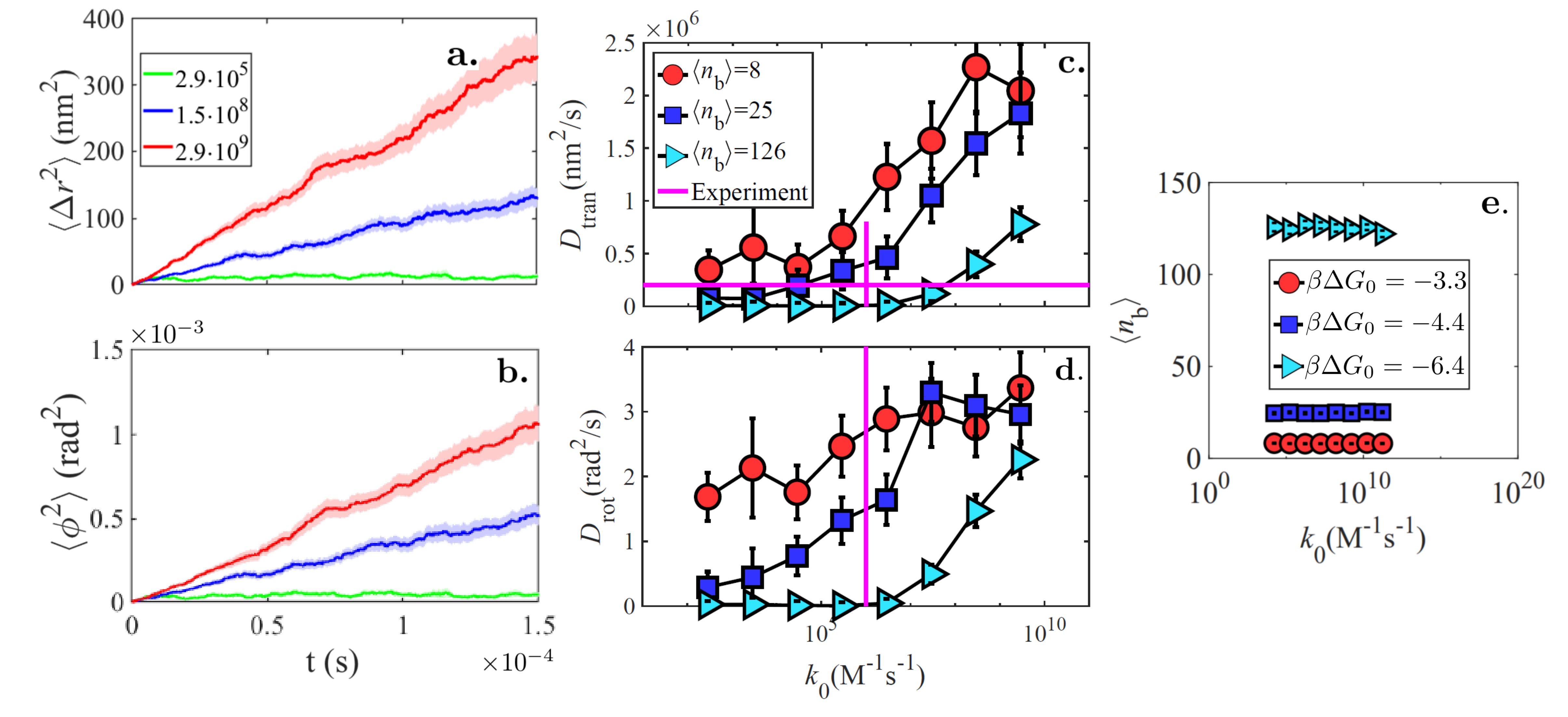} 
  \caption{{\bf Translational and rotational diffusion dynamics.}  {\bf a.}-{\bf b.} Lateral and rotational displacement (see Fig.~\ref{Fig:Fig1}) as a function of time for three different reaction rates $k_0$ (see Eq.~\ref{Equation1}) and $\langle n_b \rangle=126$. The shaded regions represent the statistical errors calculated using 50 trajectories \cite{jana2017irreversibility,PhysRevE.98.062607}. {\bf c.}-{\bf d.} Translational and rotational diffusion constants as a function of $k_0$ for $\langle n_b\rangle=8,25,126$ corresponding to $\beta \Delta G_0=-6.4,-4.4,-3.3$, respectively. In all these calculations $N_b= 5\cdot  10^4$. The vertical line corresponds to the typical binding rate of DNA sticky ends free in solution. The horizontal line in {\bf c.} represents the typical value of $D_\mathrm{tran}$ at the melting transition measured in Ref.~\cite{wang2015crystallization}. {\bf e.} At a given value of $\Delta G_0$, different values of $k_0$ do not alter the number of linkages. }\label{Fig:Fig2}
\end{figure*}

{\bf Analysis.} We define by ${\bf e}^A_\alpha(t)$ and ${\bf e}^{B}_\alpha(t)$ ($\alpha=1,\,2,\,3$) the internal reference frame, respectively, of particle $A$ and $B$. To study the relative motion between colloids, we consider the dynamics of the internal reference frame of particle $B$ in the reference frame of particle $A$, ${\bf e}^{B,A}_\alpha(t)$, ${\bf e}^{B,A}_\alpha(t)={\bf O} {\bf e}^{B}_\alpha(t)$ (see Fig.~\ref{Fig:Fig1}{\bf c}). ${\bf O}$ is a rotation matrix with line $i$ given by the transpose of ${\bf e}_{i}^A(t)$. We then calculate the translational diffusion constant, $D_\mathrm{tran}$, by studying the lateral displacement of the center of mass of $B$ ($\Delta {\bf r}$ in Fig.~\ref{Fig:Fig1}{\bf c}). Instead, we calculate the rotational diffusion constant $D_\mathrm{rot}$ from $\boldsymbol{ \phi }(t)$ defined as $\boldsymbol{\phi}(t)=\sum_{\overline t<t} {\bf w} (\overline t)$ with ${\bf w} (\overline t)= {\bf e}^{B,A}_\alpha(\overline t) \times  {\bf e}^{B,A}_\alpha(\overline t+\Delta t)$ at a given $\alpha$ \cite{hunter2011tracking}. To quantify sliding {\em versus} rolling motion, we also study the lateral displacement of the closest, $c$, and furthermost, $f$, points of B from A ($\Delta {\bf r}_c$ and $\Delta {\bf r}_f$ in Fig.~\ref{Fig:Fig1}{\bf c}). We define the translational and rotational diffusion coefficients as $\langle\Delta r^2\rangle=4D_\mathrm{tran}t$ and $\langle\phi^2\rangle=4D_\mathrm{rot}t$, respectively.

{\bf Results and Discussion.} 
In Fig.~\ref{Fig:Fig2}{\bf a}-{\bf d}, we study $D_\mathrm{tran}$ and $D_\mathrm{rot}$ as a function of the reaction rate, $k_0$, and the average number of linkages, $\langle n_b \rangle$. As predicted by Eq.~\ref{Equation1}, different values of $k_0$ do not alter the binding free energy, $\Delta G^\mathrm{hyb}_{ij}=-k_B T \log(k^\mathrm{on}_{ij}/k^\mathrm{off})$, and therefore the  average number of linkages $\langle n_b \rangle$ (Fig.~\ref{Fig:Fig2}{\bf e}). In Fig.~\ref{Fig:Fig2}{\bf a} and {\bf b} we focus, respectively, on $\langle \Delta r^2 \rangle$ and $\langle \phi^2 \rangle$ at three different values of $k_0$ for systems with $\langle n_b \rangle=126$. We find a diffusive trend for the two highest values of $k_0$ while for the smallest one the dynamics are arrested. Fig.~\ref{Fig:Fig2}{\bf c} and \ref{Fig:Fig2}{\bf d} reports, respectively, $D_\mathrm{tran}$ and $D_\mathrm{rot}$ for three different values of $\langle n_b \rangle$ (corresponding to three different values of $\Delta G_0$) and $k_0$ spanning 10 orders of magnitudes. 
Error bars a calculated as in Refs.~\cite{jana2017irreversibility,PhysRevE.98.062607}. 
Overall the diffusion coefficients increase with $k_0$ and decrease with $\langle n_b \rangle$. We observe arrested dynamics only when $\langle n_b \rangle=126$ and $k_0<10^8$M$^{-1}$s$^{-1}$. Our results show how stickiness between particles arises from having many persistent linkages. Instead, when linkages are often reconfigured (at high $k_0$), the interfaces can move past each other even when $\langle n_b \rangle > 0$. Below, we clarify how the linkages couple the rotational and translational motion of the particles resulting in rolling dynamics.  \\
The phenomenology of Fig.~\ref{Fig:Fig2} is consistent with the experimental results of Ref.~\cite{wang2015crystallization} studying the relative motion of two bound colloids (one of which is kept fixed) at the melting transition. Well below the melting temperature (corresponding to high values of $\langle n_b \rangle$) the motion is subdiffusive. Instead, close to the melting temperature (small values of $\langle n_b \rangle$) Ref. \cite{wang2015crystallization} reported diffusive motion with $D_\mathrm{tran}$ comparable to the horizontal magenta line in Fig.~\ref{Fig:Fig2}{\bf a} (corresponding to $D_\mathrm{tran}=0.2 \mu$m$^2$s$^{-1}$). A sound estimation of $k_0$ for short DNA oligomers is $k_0=10^6$M$^{-1}$s$^{-1}$ \cite{zhang2009control} (see the vertical line in Fig.~\ref{Fig:Fig2}{\bf c} and \ref{Fig:Fig2}{\bf d}). The experimental value of $D_\mathrm{tran}$ corresponds to systems with around $10^1$ linkages as usually found at the melting transition \cite{A-UbertiPCCP2016}. The successful comparison between simulations and experiments corroborates our model that will be useful to study the rotational motion of the particles which cannot be studied in experiments (see Fig.~\ref{Fig:Fig2}{\bf d} and \ref{Fig:Fig4}).
\begin{figure}
\centering
\includegraphics[scale=0.38]{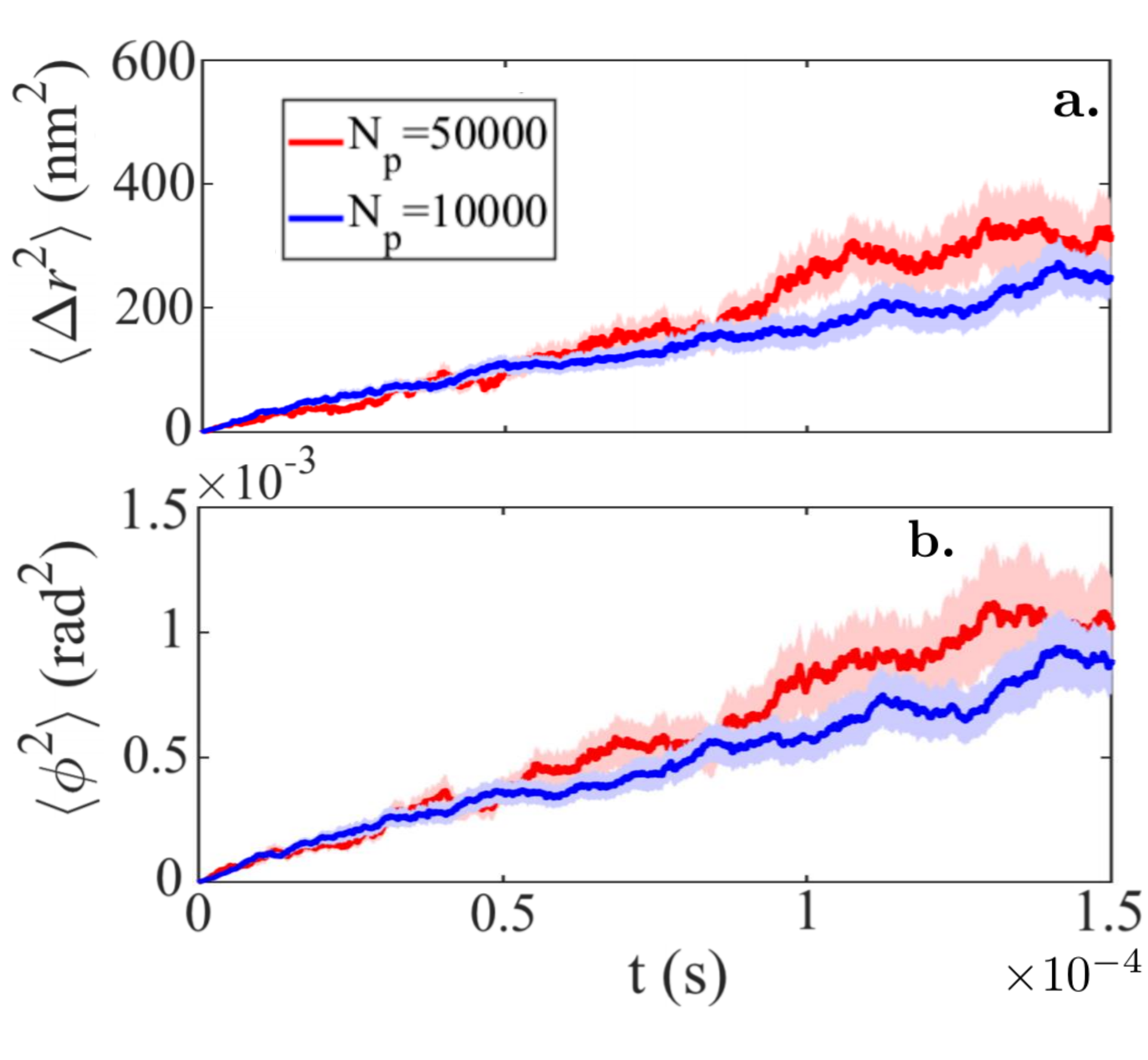}
\caption{{\bf The rate at which linkages break, $k^\mathrm{off}$, controls the diffusivity of cross-linked interfaces.} {\bf a.}-{\bf b.} Systems with the same number of linkages, $\langle n_b\rangle$, and same $off$ rate, $k^\mathrm{off}$, feature the same translational and rotational displacement. We use $\langle n_b \rangle = 26-29$ and $k^\mathrm{off}=3.6\cdot 10^4$s$^{-1}$ corresponding to $k_0=2.9\cdot 10^6$M$^{-1}$s$^{-1}$ and $k_0=55\cdot2.9\cdot 10^6$M$^{-1}$s$^{-1}$, respectively, for the system with $N_p=10^4$ and $N_p=5\cdot 10^4$ linkers (see Eq.~\ref{Equation1}). The statistical analysis is based on 50 independent trajectories.
}\label{Fig:Fig3}
\end{figure}
\\
The results of Fig. \ref{Fig:Fig2} prove the pivotal role played by $k_0$, along with the number of bridges $\langle n_b \rangle$, in controlling the dynamics of the system. We now clarify how the single bridge lifetime, $(k^\mathrm{off})^{-1}$, is the only timescale of the model that affects colloids’ diffusion (at a given $\langle n_b \rangle$). To do so, in Fig.~\ref{Fig:Fig3} we compare two systems with a total number of linkers {\em per} particle equal to $N_p= 10^4$ and $N_p=5\cdot 10^4$. 
These two systems feature coating densities comparable with the ones found in two classes of experiments employing different tethering molecules, Ref.~\cite{RogersPNAS2011} and Ref.~\cite{wang2015crystallization,wang2015synthetic}. We fine-tune the values of $\Delta G_0$ in such a way that the numbers of bridges displayed by the two pairs of particles remain comparable, $\langle n_b \rangle \approx 26-29$. In particular, for the $N_p= 10^4$ and $N_p=5\cdot 10^4$ case, we use, respectively,  $\beta \Delta G_0=-8.4$ and $\beta \Delta G_0=-4.4$. As expected, bridges are stronger ($\Delta G_0$ is lower) for $N_p=10^4$ than for $N_p=5\cdot 10^4$ due to combinatorial entropy terms favoring bridge formation in the latter case. Given the two values of $\Delta G_0$, we calculate the corresponding values of $k_0$ (see caption of Fig. \ref{Fig:Fig3}) leading to the same $k^\mathrm{off}$ as predicted by Eq.~\ref{Equation1}. We then simulate the diffusion dynamics of the two systems (Fig. \ref{Fig:Fig3}). Remarkably, Fig. \ref{Fig:Fig3} shows how the two systems feature identical rotational and translational dynamics. 


\begin{figure}
\centering
  \includegraphics[scale=0.27]{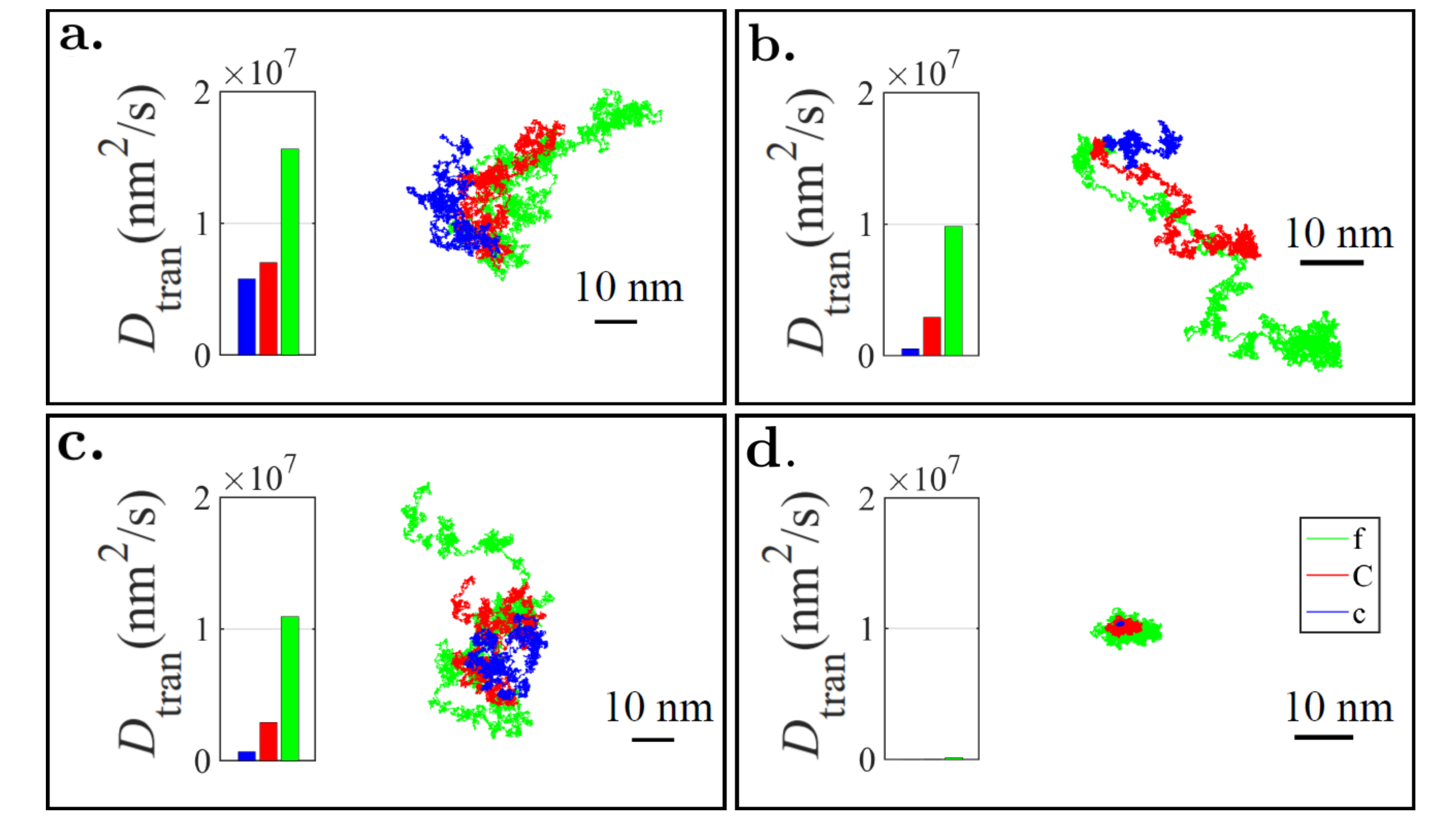}
  \vspace{0.01cm}
  \caption{{\bf The internal dynamics of crosslinked colloids.} While changing the number of linkages, $\langle n_b \rangle$, and the kinetic rates, $k_0$, colloids slide {\bf a}, roll ({\bf b} and {\bf c}), or stop diffusing ({\bf d}). Particles roll when the translational diffusion of a reference point in the contact region, $c$, becomes much smaller than the diffusion of the center of mass, $C$, or $f$ (see Fig.~\ref{Fig:Fig1}{\bf c} for the definitions). We calculate $D_\mathrm{tran}$ by averaging over 50 trajectories. We also report single trajectories of $\Delta {\bf r}_c$ ($c$), $\Delta {\bf r}$ ($C$), and $\Delta {\bf r}_f$ (see Fig.~\ref{Fig:Fig1}{\bf c}). The employed values of $\langle n_b \rangle$ and $k_0$ are: {\bf a.} 
  $k_0=2.9\cdot 10^9$M$^{-1}$s$^{-1}$ and $\langle n_b\rangle=8$, {\bf b.} $k_0=2.9\cdot 10^9$ M$^{-1}$s$^{-1}$ and $\langle n_b\rangle=126$, {\bf c.} $k_0=2.9\cdot 10^5$M$^{-1}$s$^{-1}$ and $\langle n_b\rangle=8$, {\bf d.} $k_0=2.9\cdot 10^5$M$^{-1}$s$^{-1}$ and $\langle n_b\rangle=126$.}
  \label{Fig:Fig4}
\end{figure}

We now study how the linkages constrain the relative dynamics of the two particles resulting in a coupling between the rotational and the translational motion of the colloids. In turn, this coupling enhances rolling dynamics over sliding \cite{C8SM01430B}. A colloid slides if its internal reference frame remains parallel to the tangential plane of the opposing surface, while rolls if the closest point to the facing surface, $c$ in Fig.~\ref{Fig:Fig1}{\bf c}, remains at rest during an update. Therefore in Fig.~\ref{Fig:Fig4} we compare the lateral displacement of point $c$ with the one of point $f$ and the center of mass $C$ (see Fig.~\ref{Fig:Fig1}{\bf c}). Because the position of the contact point does not change significantly during our simulations, we fix $c$ (and similarly $f$) to the contact point at time $t=0$ (see Fig.~\ref{Fig:Fig1}{\bf c}). 
The diffusion of the contact point $c$ is maximized at low values of $\langle n_b \rangle$ and high reaction rates, $k_0$, (see Fig.~\ref{Fig:Fig4}{\bf a}). In such conditions, colloids slide. When increasing the number of bridges (Fig.~\ref{Fig:Fig4}{\bf b}) or decreasing the reaction rate (Fig.~\ref{Fig:Fig4}{\bf c}) the diffusion of the point $c$ drastically slow down, and colloids start rolling. Intriguingly, the diffusion constants of Figs.~\ref{Fig:Fig4}{\bf b} and \ref{Fig:Fig4}{\bf c} are comparable. Finally, at low $k_0$ and high $\langle n_b \rangle$, the dynamics are arrested (Fig.~\ref{Fig:Fig4}{\bf d}). A recent theoretical contribution also predicted rolling motion \cite{C8SM01430B}. Unfortunately, Ref.~\cite{C8SM01430B} used a simplified model in which linkers bind any point of the facing surface (instead of complementary partners) hampering a direct comparison with our results.

{\bf Conclusions.} In this work we have studied the internal dynamics of colloids decorated by linkers with fixed tethering points forming reversible inter-particle linkages. We have clarified how the average number of linkages, $\langle n_b \rangle$, and the rates at which linkages form and break, $k_0$ in Eq.~\ref{Equation1}, are the key parameters controlling the relative diffusion of cross-linked particles. We confirm the results of recent experiments employing DNA linkers showing how cross-linked colloids can diffuse while being bound \cite{wang2015crystallization}. We explain how this result is due to the higher coating density employed by Ref.~\cite{wang2015crystallization} than what used in previous studies resulting in faster reaction kinetics at the melting transition. We also clarify how the balance between the number of linkages $\langle n_b \rangle$ and reaction rates $k_0$ leads to rolling motions.  Our results and quantitative methodology will be useful to study biological systems, like virus infection, where the number of linkages between host and pathogen is tightly regulated (e.g.~\cite{Delgusteeaat1273}).

{\bf Acknowledgement.} 
This work was supported by the Fonds de la Recherche
Scientifique de Belgique (F.R.S.)-FNRS under grant n$^\circ$ MIS
F.4534.17. Computational resources have been provided
by the {\em Consortium des \'Equipements de Calcul Intensif} 
(CECI), funded by the F.R.S.-FNRS under grant n$^\circ$ 2.5020.11.


\begin{thebibliography}{58}%
\makeatletter
\providecommand \@ifxundefined [1]{%
 \@ifx{#1\undefined}
}%
\providecommand \@ifnum [1]{%
 \ifnum #1\expandafter \@firstoftwo
 \else \expandafter \@secondoftwo
 \fi
}%
\providecommand \@ifx [1]{%
 \ifx #1\expandafter \@firstoftwo
 \else \expandafter \@secondoftwo
 \fi
}%
\providecommand \natexlab [1]{#1}%
\providecommand \enquote  [1]{``#1''}%
\providecommand \bibnamefont  [1]{#1}%
\providecommand \bibfnamefont [1]{#1}%
\providecommand \citenamefont [1]{#1}%
\providecommand \href@noop [0]{\@secondoftwo}%
\providecommand \href [0]{\begingroup \@sanitize@url \@href}%
\providecommand \@href[1]{\@@startlink{#1}\@@href}%
\providecommand \@@href[1]{\endgroup#1\@@endlink}%
\providecommand \@sanitize@url [0]{\catcode `\\12\catcode `\$12\catcode
  `\&12\catcode `\#12\catcode `\^12\catcode `\_12\catcode `\%12\relax}%
\providecommand \@@startlink[1]{}%
\providecommand \@@endlink[0]{}%
\providecommand \url  [0]{\begingroup\@sanitize@url \@url }%
\providecommand \@url [1]{\endgroup\@href {#1}{\urlprefix }}%
\providecommand \urlprefix  [0]{URL }%
\providecommand \Eprint [0]{\href }%
\providecommand \doibase [0]{http://dx.doi.org/}%
\providecommand \selectlanguage [0]{\@gobble}%
\providecommand \bibinfo  [0]{\@secondoftwo}%
\providecommand \bibfield  [0]{\@secondoftwo}%
\providecommand \translation [1]{[#1]}%
\providecommand \BibitemOpen [0]{}%
\providecommand \bibitemStop [0]{}%
\providecommand \bibitemNoStop [0]{.\EOS\space}%
\providecommand \EOS [0]{\spacefactor3000\relax}%
\providecommand \BibitemShut  [1]{\csname bibitem#1\endcsname}%
\let\auto@bib@innerbib\@empty
\bibitem [{\citenamefont {Park}\ \emph {et~al.}(2008)\citenamefont {Park},
  \citenamefont {Lytton-Jean}, \citenamefont {Lee}, \citenamefont {Weigand},
  \citenamefont {Schatz},\ and\ \citenamefont {Mirkin}}]{park2008dna}%
  \BibitemOpen
  \bibfield  {author} {\bibinfo {author} {\bibfnamefont {S.~Y.}\ \bibnamefont
  {Park}}, \bibinfo {author} {\bibfnamefont {A.~K.}\ \bibnamefont
  {Lytton-Jean}}, \bibinfo {author} {\bibfnamefont {B.}~\bibnamefont {Lee}},
  \bibinfo {author} {\bibfnamefont {S.}~\bibnamefont {Weigand}}, \bibinfo
  {author} {\bibfnamefont {G.~C.}\ \bibnamefont {Schatz}}, \ and\ \bibinfo
  {author} {\bibfnamefont {C.~A.}\ \bibnamefont {Mirkin}},\ }\href@noop {}
  {\bibfield  {journal} {\bibinfo  {journal} {Nature}\ }\textbf {\bibinfo
  {volume} {451}},\ \bibinfo {pages} {553} (\bibinfo {year}
  {2008})}\BibitemShut {NoStop}%
\bibitem [{\citenamefont {Auyeung}\ \emph {et~al.}(2014)\citenamefont
  {Auyeung}, \citenamefont {Li}, \citenamefont {Senesi}, \citenamefont
  {Schmucker}, \citenamefont {Pals}, \citenamefont {Olvera~de La~Cruz},\ and\
  \citenamefont {Mirkin}}]{auyeung2014dna}%
  \BibitemOpen
  \bibfield  {author} {\bibinfo {author} {\bibfnamefont {E.}~\bibnamefont
  {Auyeung}}, \bibinfo {author} {\bibfnamefont {T.~I.}\ \bibnamefont {Li}},
  \bibinfo {author} {\bibfnamefont {A.~J.}\ \bibnamefont {Senesi}}, \bibinfo
  {author} {\bibfnamefont {A.~L.}\ \bibnamefont {Schmucker}}, \bibinfo {author}
  {\bibfnamefont {B.~C.}\ \bibnamefont {Pals}}, \bibinfo {author}
  {\bibfnamefont {M.}~\bibnamefont {Olvera~de La~Cruz}}, \ and\ \bibinfo
  {author} {\bibfnamefont {C.~A.}\ \bibnamefont {Mirkin}},\ }\href@noop {}
  {\bibfield  {journal} {\bibinfo  {journal} {Nature}\ }\textbf {\bibinfo
  {volume} {505}},\ \bibinfo {pages} {73} (\bibinfo {year} {2014})}\BibitemShut
  {NoStop}%
\bibitem [{\citenamefont {Rogers}\ and\ \citenamefont
  {Manoharan}(2015)}]{rogers-manoharan}%
  \BibitemOpen
  \bibfield  {author} {\bibinfo {author} {\bibfnamefont {W.~B.}\ \bibnamefont
  {Rogers}}\ and\ \bibinfo {author} {\bibfnamefont {V.~N.}\ \bibnamefont
  {Manoharan}},\ }\href@noop {} {\bibfield  {journal} {\bibinfo  {journal}
  {Science}\ }\textbf {\bibinfo {volume} {347}},\ \bibinfo {pages} {639}
  (\bibinfo {year} {2015})}\BibitemShut {NoStop}%
\bibitem [{\citenamefont {Ducrot}\ \emph {et~al.}(2017)\citenamefont {Ducrot},
  \citenamefont {He}, \citenamefont {Yi},\ and\ \citenamefont
  {Pine}}]{Ducrot_NatMat_2017}%
  \BibitemOpen
  \bibfield  {author} {\bibinfo {author} {\bibfnamefont {{\'E}.}~\bibnamefont
  {Ducrot}}, \bibinfo {author} {\bibfnamefont {M.}~\bibnamefont {He}}, \bibinfo
  {author} {\bibfnamefont {G.-R.}\ \bibnamefont {Yi}}, \ and\ \bibinfo {author}
  {\bibfnamefont {D.~J.}\ \bibnamefont {Pine}},\ }\href@noop {} {\bibfield
  {journal} {\bibinfo  {journal} {Nat.\ Mater.}\ }\textbf {\bibinfo {volume}
  {16}},\ \bibinfo {pages} {652} (\bibinfo {year} {2017})}\BibitemShut
  {NoStop}%
\bibitem [{\citenamefont {Wang}\ \emph {et~al.}(2017)\citenamefont {Wang},
  \citenamefont {Jenkins}, \citenamefont {McGinley}, \citenamefont {Sinno},\
  and\ \citenamefont {Crocker}}]{Wang_NatComm_2017}%
  \BibitemOpen
  \bibfield  {author} {\bibinfo {author} {\bibfnamefont {Y.}~\bibnamefont
  {Wang}}, \bibinfo {author} {\bibfnamefont {I.~C.}\ \bibnamefont {Jenkins}},
  \bibinfo {author} {\bibfnamefont {J.~T.}\ \bibnamefont {McGinley}}, \bibinfo
  {author} {\bibfnamefont {T.}~\bibnamefont {Sinno}}, \ and\ \bibinfo {author}
  {\bibfnamefont {J.~C.}\ \bibnamefont {Crocker}},\ }\href@noop {} {\bibfield
  {journal} {\bibinfo  {journal} {Nat.\ Commun.}\ }\textbf {\bibinfo {volume}
  {8}},\ \bibinfo {pages} {14173} (\bibinfo {year} {2017})}\BibitemShut
  {NoStop}%
\bibitem [{\citenamefont {Liu}\ \emph {et~al.}(2016)\citenamefont {Liu},
  \citenamefont {Tagawa}, \citenamefont {Xin}, \citenamefont {Wang},
  \citenamefont {Emamy}, \citenamefont {Li}, \citenamefont {Yager},
  \citenamefont {Starr}, \citenamefont {Tkachenko},\ and\ \citenamefont
  {Gang}}]{Liu_Science_2016}%
  \BibitemOpen
  \bibfield  {author} {\bibinfo {author} {\bibfnamefont {W.}~\bibnamefont
  {Liu}}, \bibinfo {author} {\bibfnamefont {M.}~\bibnamefont {Tagawa}},
  \bibinfo {author} {\bibfnamefont {H.~L.}\ \bibnamefont {Xin}}, \bibinfo
  {author} {\bibfnamefont {T.}~\bibnamefont {Wang}}, \bibinfo {author}
  {\bibfnamefont {H.}~\bibnamefont {Emamy}}, \bibinfo {author} {\bibfnamefont
  {H.}~\bibnamefont {Li}}, \bibinfo {author} {\bibfnamefont {K.~G.}\
  \bibnamefont {Yager}}, \bibinfo {author} {\bibfnamefont {F.~W.}\ \bibnamefont
  {Starr}}, \bibinfo {author} {\bibfnamefont {A.~V.}\ \bibnamefont
  {Tkachenko}}, \ and\ \bibinfo {author} {\bibfnamefont {O.}~\bibnamefont
  {Gang}},\ }\href {\doibase 10.1126/science.aad2080} {\bibfield  {journal}
  {\bibinfo  {journal} {Science}\ }\textbf {\bibinfo {volume} {351}},\ \bibinfo
  {pages} {582} (\bibinfo {year} {2016})}\BibitemShut {NoStop}%
\bibitem [{\citenamefont {Biffi}\ \emph {et~al.}(2013)\citenamefont {Biffi},
  \citenamefont {Cerbino}, \citenamefont {Bomboi}, \citenamefont {Paraboschi},
  \citenamefont {Asselta}, \citenamefont {Sciortino},\ and\ \citenamefont
  {Bellini}}]{Biffi_PNAS_2013}%
  \BibitemOpen
  \bibfield  {author} {\bibinfo {author} {\bibfnamefont {S.}~\bibnamefont
  {Biffi}}, \bibinfo {author} {\bibfnamefont {R.}~\bibnamefont {Cerbino}},
  \bibinfo {author} {\bibfnamefont {F.}~\bibnamefont {Bomboi}}, \bibinfo
  {author} {\bibfnamefont {E.~M.}\ \bibnamefont {Paraboschi}}, \bibinfo
  {author} {\bibfnamefont {R.}~\bibnamefont {Asselta}}, \bibinfo {author}
  {\bibfnamefont {F.}~\bibnamefont {Sciortino}}, \ and\ \bibinfo {author}
  {\bibfnamefont {T.}~\bibnamefont {Bellini}},\ }\href@noop {} {\bibfield
  {journal} {\bibinfo  {journal} {Proc.\ Natl.\ Acad.\ Sci.\ U.\ S.\ A.}\
  }\textbf {\bibinfo {volume} {110}},\ \bibinfo {pages} {15633} (\bibinfo
  {year} {2013})}\BibitemShut {NoStop}%
\bibitem [{\citenamefont {Di~Michele}\ \emph {et~al.}(2014)\citenamefont
  {Di~Michele}, \citenamefont {Fiocco}, \citenamefont {Varrato}, \citenamefont
  {Sastry}, \citenamefont {Eiser},\ and\ \citenamefont
  {Foffi}}]{di2014aggregation}%
  \BibitemOpen
  \bibfield  {author} {\bibinfo {author} {\bibfnamefont {L.}~\bibnamefont
  {Di~Michele}}, \bibinfo {author} {\bibfnamefont {D.}~\bibnamefont {Fiocco}},
  \bibinfo {author} {\bibfnamefont {F.}~\bibnamefont {Varrato}}, \bibinfo
  {author} {\bibfnamefont {S.}~\bibnamefont {Sastry}}, \bibinfo {author}
  {\bibfnamefont {E.}~\bibnamefont {Eiser}}, \ and\ \bibinfo {author}
  {\bibfnamefont {G.}~\bibnamefont {Foffi}},\ }\href@noop {} {\bibfield
  {journal} {\bibinfo  {journal} {Soft Matter}\ }\textbf {\bibinfo {volume}
  {10}},\ \bibinfo {pages} {3633} (\bibinfo {year} {2014})}\BibitemShut
  {NoStop}%
\bibitem [{\citenamefont {McMullen}\ \emph {et~al.}(2018)\citenamefont
  {McMullen}, \citenamefont {Holmes-Cerfon}, \citenamefont {Sciortino},
  \citenamefont {Grosberg},\ and\ \citenamefont {Brujic}}]{Jasna_PRL_2018}%
  \BibitemOpen
  \bibfield  {author} {\bibinfo {author} {\bibfnamefont {A.}~\bibnamefont
  {McMullen}}, \bibinfo {author} {\bibfnamefont {M.}~\bibnamefont
  {Holmes-Cerfon}}, \bibinfo {author} {\bibfnamefont {F.}~\bibnamefont
  {Sciortino}}, \bibinfo {author} {\bibfnamefont {A.~Y.}\ \bibnamefont
  {Grosberg}}, \ and\ \bibinfo {author} {\bibfnamefont {J.}~\bibnamefont
  {Brujic}},\ }\href {\doibase 10.1103/PhysRevLett.121.138002} {\bibfield
  {journal} {\bibinfo  {journal} {Phys. Rev. Lett.}\ }\textbf {\bibinfo
  {volume} {121}},\ \bibinfo {pages} {138002} (\bibinfo {year}
  {2018})}\BibitemShut {NoStop}%
\bibitem [{\citenamefont {Alberts}\ \emph {et~al.}(2015)\citenamefont
  {Alberts}, \citenamefont {Johnson}, \citenamefont {Lewis}, \citenamefont
  {Morgan}, \citenamefont {Raff}, \citenamefont {Roberts},\ and\ \citenamefont
  {Walter}}]{Alberts}%
  \BibitemOpen
  \bibfield  {author} {\bibinfo {author} {\bibfnamefont {B.}~\bibnamefont
  {Alberts}}, \bibinfo {author} {\bibfnamefont {A.}~\bibnamefont {Johnson}},
  \bibinfo {author} {\bibfnamefont {J.}~\bibnamefont {Lewis}}, \bibinfo
  {author} {\bibfnamefont {D.}~\bibnamefont {Morgan}}, \bibinfo {author}
  {\bibfnamefont {M.}~\bibnamefont {Raff}}, \bibinfo {author} {\bibfnamefont
  {K.}~\bibnamefont {Roberts}}, \ and\ \bibinfo {author} {\bibfnamefont
  {P.}~\bibnamefont {Walter}},\ }\href@noop {} {\emph {\bibinfo {title}
  {Molecular Biology of the Cell, Sixth Edition}}}\ (\bibinfo  {publisher}
  {Garland Science},\ \bibinfo {year} {2015})\BibitemShut {NoStop}%
\bibitem [{\citenamefont {Knorowski}\ and\ \citenamefont
  {Travesset}(2011)}]{knorowski2011materials}%
  \BibitemOpen
  \bibfield  {author} {\bibinfo {author} {\bibfnamefont {C.}~\bibnamefont
  {Knorowski}}\ and\ \bibinfo {author} {\bibfnamefont {A.}~\bibnamefont
  {Travesset}},\ }\href@noop {} {\bibfield  {journal} {\bibinfo  {journal}
  {Curr. Opin. Solid St. M. Sci.}\ }\textbf {\bibinfo {volume} {15}},\ \bibinfo
  {pages} {262} (\bibinfo {year} {2011})}\BibitemShut {NoStop}%
\bibitem [{\citenamefont {Angioletti-Uberti}\ \emph {et~al.}(2016)\citenamefont
  {Angioletti-Uberti}, \citenamefont {Mognetti},\ and\ \citenamefont
  {Frenkel}}]{A-UbertiPCCP2016}%
  \BibitemOpen
  \bibfield  {author} {\bibinfo {author} {\bibfnamefont {S.}~\bibnamefont
  {Angioletti-Uberti}}, \bibinfo {author} {\bibfnamefont {B.~M.}\ \bibnamefont
  {Mognetti}}, \ and\ \bibinfo {author} {\bibfnamefont {D.}~\bibnamefont
  {Frenkel}},\ }\href {\doibase 10.1039/C5CP06981E} {\bibfield  {journal}
  {\bibinfo  {journal} {Phys. Chem. Chem. Phys.}\ }\textbf {\bibinfo {volume}
  {18}},\ \bibinfo {pages} {6373} (\bibinfo {year} {2016})}\BibitemShut
  {NoStop}%
\bibitem [{\citenamefont {Bell}\ \emph {et~al.}(1984)\citenamefont {Bell},
  \citenamefont {Dembo},\ and\ \citenamefont {Bongrand}}]{BellBJourn1984}%
  \BibitemOpen
  \bibfield  {author} {\bibinfo {author} {\bibfnamefont {G.~I.}\ \bibnamefont
  {Bell}}, \bibinfo {author} {\bibfnamefont {M.}~\bibnamefont {Dembo}}, \ and\
  \bibinfo {author} {\bibfnamefont {P.}~\bibnamefont {Bongrand}},\ }\href@noop
  {} {\bibfield  {journal} {\bibinfo  {journal} {Biophys. J.}\ }\textbf
  {\bibinfo {volume} {45}},\ \bibinfo {pages} {1051} (\bibinfo {year}
  {1984})}\BibitemShut {NoStop}%
\bibitem [{\citenamefont {Dreyfus}\ \emph {et~al.}(2009)\citenamefont
  {Dreyfus}, \citenamefont {Leunissen}, \citenamefont {Sha}, \citenamefont
  {Tkachenko}, \citenamefont {Seeman}, \citenamefont {Pine},\ and\
  \citenamefont {Chaikin}}]{melting-theory1}%
  \BibitemOpen
  \bibfield  {author} {\bibinfo {author} {\bibfnamefont {R.}~\bibnamefont
  {Dreyfus}}, \bibinfo {author} {\bibfnamefont {M.~E.}\ \bibnamefont
  {Leunissen}}, \bibinfo {author} {\bibfnamefont {R.}~\bibnamefont {Sha}},
  \bibinfo {author} {\bibfnamefont {A.~V.}\ \bibnamefont {Tkachenko}}, \bibinfo
  {author} {\bibfnamefont {N.~C.}\ \bibnamefont {Seeman}}, \bibinfo {author}
  {\bibfnamefont {D.~J.}\ \bibnamefont {Pine}}, \ and\ \bibinfo {author}
  {\bibfnamefont {P.~M.}\ \bibnamefont {Chaikin}},\ }\href {\doibase
  10.1103/PhysRevLett.102.048301} {\bibfield  {journal} {\bibinfo  {journal}
  {Phys. Rev. Lett.}\ }\textbf {\bibinfo {volume} {102}},\ \bibinfo {pages}
  {048301} (\bibinfo {year} {2009})}\BibitemShut {NoStop}%
\bibitem [{\citenamefont {Rogers}\ and\ \citenamefont
  {Crocker}(2011)}]{RogersPNAS2011}%
  \BibitemOpen
  \bibfield  {author} {\bibinfo {author} {\bibfnamefont {W.~B.}\ \bibnamefont
  {Rogers}}\ and\ \bibinfo {author} {\bibfnamefont {J.~C.}\ \bibnamefont
  {Crocker}},\ }\href {\doibase 10.1073/pnas.1109853108} {\bibfield  {journal}
  {\bibinfo  {journal} {Proc. Natl. Acad. Sci. USA}\ }\textbf {\bibinfo
  {volume} {108}},\ \bibinfo {pages} {15687} (\bibinfo {year} {2011})},\
  \Eprint
  {http://arxiv.org/abs/http://www.pnas.org/content/108/38/15687.full.pdf+html}
  {http://www.pnas.org/content/108/38/15687.full.pdf+html} \BibitemShut
  {NoStop}%
\bibitem [{\citenamefont {Mladek}\ \emph {et~al.}(2012)\citenamefont {Mladek},
  \citenamefont {Fornleitner}, \citenamefont {Martinez-Veracoechea},
  \citenamefont {Dawid},\ and\ \citenamefont {Frenkel}}]{MladekPRL2012}%
  \BibitemOpen
  \bibfield  {author} {\bibinfo {author} {\bibfnamefont {B.~M.}\ \bibnamefont
  {Mladek}}, \bibinfo {author} {\bibfnamefont {J.}~\bibnamefont {Fornleitner}},
  \bibinfo {author} {\bibfnamefont {F.~J.}\ \bibnamefont
  {Martinez-Veracoechea}}, \bibinfo {author} {\bibfnamefont {A.}~\bibnamefont
  {Dawid}}, \ and\ \bibinfo {author} {\bibfnamefont {D.}~\bibnamefont
  {Frenkel}},\ }\href@noop {} {\bibfield  {journal} {\bibinfo  {journal} {Phys.
  Rev. Lett.}\ }\textbf {\bibinfo {volume} {108}},\ \bibinfo {pages} {268301}
  (\bibinfo {year} {2012})}\BibitemShut {NoStop}%
\bibitem [{\citenamefont {Martinez-Veracoechea}\ and\ \citenamefont
  {Frenkel}(2011)}]{francisco-pnas}%
  \BibitemOpen
  \bibfield  {author} {\bibinfo {author} {\bibfnamefont {F.~J.}\ \bibnamefont
  {Martinez-Veracoechea}}\ and\ \bibinfo {author} {\bibfnamefont
  {D.}~\bibnamefont {Frenkel}},\ }\href {\doibase 10.1073/pnas.1105351108}
  {\bibfield  {journal} {\bibinfo  {journal} {Proc. Natl. Acad. Sci. USA}\
  }\textbf {\bibinfo {volume} {108}},\ \bibinfo {pages} {10963} (\bibinfo
  {year} {2011})}\BibitemShut {NoStop}%
\bibitem [{\citenamefont {Tito}\ \emph {et~al.}(2016)\citenamefont {Tito},
  \citenamefont {Angioletti-Uberti},\ and\ \citenamefont
  {Frenkel}}]{TitoJCP2016}%
  \BibitemOpen
  \bibfield  {author} {\bibinfo {author} {\bibfnamefont {N.~B.}\ \bibnamefont
  {Tito}}, \bibinfo {author} {\bibfnamefont {S.}~\bibnamefont
  {Angioletti-Uberti}}, \ and\ \bibinfo {author} {\bibfnamefont
  {D.}~\bibnamefont {Frenkel}},\ }\href@noop {} {\bibfield  {journal} {\bibinfo
   {journal} {J. Chem. Phys.}\ }\textbf {\bibinfo {volume} {144}},\ \bibinfo
  {pages} {161101} (\bibinfo {year} {2016})}\BibitemShut {NoStop}%
\bibitem [{\citenamefont {Zhdanov}(2017)}]{PhysRevE.96.012408}%
  \BibitemOpen
  \bibfield  {author} {\bibinfo {author} {\bibfnamefont {V.~P.}\ \bibnamefont
  {Zhdanov}},\ }\href {\doibase 10.1103/PhysRevE.96.012408} {\bibfield
  {journal} {\bibinfo  {journal} {Phys. Rev. E}\ }\textbf {\bibinfo {volume}
  {96}},\ \bibinfo {pages} {012408} (\bibinfo {year} {2017})}\BibitemShut
  {NoStop}%
\bibitem [{\citenamefont {Jenkins}\ \emph {et~al.}(2017)\citenamefont
  {Jenkins}, \citenamefont {Crocker},\ and\ \citenamefont
  {Sinno}}]{jenkins2017interaction}%
  \BibitemOpen
  \bibfield  {author} {\bibinfo {author} {\bibfnamefont {I.~C.}\ \bibnamefont
  {Jenkins}}, \bibinfo {author} {\bibfnamefont {J.~C.}\ \bibnamefont
  {Crocker}}, \ and\ \bibinfo {author} {\bibfnamefont {T.}~\bibnamefont
  {Sinno}},\ }\href@noop {} {\bibfield  {journal} {\bibinfo  {journal} {Phys.
  Rev. Lett.}\ }\textbf {\bibinfo {volume} {119}},\ \bibinfo {pages} {178002}
  (\bibinfo {year} {2017})}\BibitemShut {NoStop}%
\bibitem [{\citenamefont {Di~Michele}\ \emph {et~al.}(2018)\citenamefont
  {Di~Michele}, \citenamefont {Jana},\ and\ \citenamefont
  {Mognetti}}]{DiMichelePRE2018}%
  \BibitemOpen
  \bibfield  {author} {\bibinfo {author} {\bibfnamefont {L.}~\bibnamefont
  {Di~Michele}}, \bibinfo {author} {\bibfnamefont {P.~K.}\ \bibnamefont
  {Jana}}, \ and\ \bibinfo {author} {\bibfnamefont {B.~M.}\ \bibnamefont
  {Mognetti}},\ }\href {\doibase 10.1103/PhysRevE.98.032406} {\bibfield
  {journal} {\bibinfo  {journal} {Phys. Rev. E}\ }\textbf {\bibinfo {volume}
  {98}},\ \bibinfo {pages} {032406} (\bibinfo {year} {2018})}\BibitemShut
  {NoStop}%
\bibitem [{\citenamefont {Li}\ \emph {et~al.}(2012)\citenamefont {Li},
  \citenamefont {Sknepnek}, \citenamefont {Macfarlane}, \citenamefont
  {Mirkin},\ and\ \citenamefont {Olvera de~la Cruz}}]{li2012modeling}%
  \BibitemOpen
  \bibfield  {author} {\bibinfo {author} {\bibfnamefont {T.~I.}\ \bibnamefont
  {Li}}, \bibinfo {author} {\bibfnamefont {R.}~\bibnamefont {Sknepnek}},
  \bibinfo {author} {\bibfnamefont {R.~J.}\ \bibnamefont {Macfarlane}},
  \bibinfo {author} {\bibfnamefont {C.~A.}\ \bibnamefont {Mirkin}}, \ and\
  \bibinfo {author} {\bibfnamefont {M.}~\bibnamefont {Olvera de~la Cruz}},\
  }\href@noop {} {\bibfield  {journal} {\bibinfo  {journal} {Nano letters}\
  }\textbf {\bibinfo {volume} {12}},\ \bibinfo {pages} {2509} (\bibinfo {year}
  {2012})}\BibitemShut {NoStop}%
\bibitem [{\citenamefont {Mirkin}\ \emph {et~al.}(1996)\citenamefont {Mirkin},
  \citenamefont {Letsinger}, \citenamefont {Mucic},\ and\ \citenamefont
  {Storhoff}}]{mirkin}%
  \BibitemOpen
  \bibfield  {author} {\bibinfo {author} {\bibfnamefont {C.~A.}\ \bibnamefont
  {Mirkin}}, \bibinfo {author} {\bibfnamefont {R.~C.}\ \bibnamefont
  {Letsinger}}, \bibinfo {author} {\bibfnamefont {R.~C.}\ \bibnamefont
  {Mucic}}, \ and\ \bibinfo {author} {\bibfnamefont {J.~J.}\ \bibnamefont
  {Storhoff}},\ }\href {\doibase http://dx.doi.org/10.1038/382607a0} {\bibfield
   {journal} {\bibinfo  {journal} {Nature}\ }\textbf {\bibinfo {volume}
  {382}},\ \bibinfo {pages} {607} (\bibinfo {year} {1996})}\BibitemShut
  {NoStop}%
\bibitem [{\citenamefont {Alivisatos}\ \emph {et~al.}(1996)\citenamefont
  {Alivisatos}, \citenamefont {Johnsson}, \citenamefont {Peng}, \citenamefont
  {Wilson}, \citenamefont {Loweth}, \citenamefont {Bruchez},\ and\
  \citenamefont {Schultz}}]{alivisatos}%
  \BibitemOpen
  \bibfield  {author} {\bibinfo {author} {\bibfnamefont {A.~P.}\ \bibnamefont
  {Alivisatos}}, \bibinfo {author} {\bibfnamefont {K.~P.}\ \bibnamefont
  {Johnsson}}, \bibinfo {author} {\bibfnamefont {X.}~\bibnamefont {Peng}},
  \bibinfo {author} {\bibfnamefont {T.~E.}\ \bibnamefont {Wilson}}, \bibinfo
  {author} {\bibfnamefont {C.~J.}\ \bibnamefont {Loweth}}, \bibinfo {author}
  {\bibfnamefont {M.~P.}\ \bibnamefont {Bruchez}}, \ and\ \bibinfo {author}
  {\bibfnamefont {P.~G.}\ \bibnamefont {Schultz}},\ }\href {\doibase
  http://dx.doi.org/10.1038/382609a0} {\bibfield  {journal} {\bibinfo
  {journal} {Nature}\ }\textbf {\bibinfo {volume} {382}},\ \bibinfo {pages}
  {609} (\bibinfo {year} {1996})}\BibitemShut {NoStop}%
\bibitem [{\citenamefont {Dreyfus}\ \emph {et~al.}(2010)\citenamefont
  {Dreyfus}, \citenamefont {Leunissen}, \citenamefont {Sha}, \citenamefont
  {Tkachenko}, \citenamefont {Seeman}, \citenamefont {Pine},\ and\
  \citenamefont {Chaikin}}]{melting-theory2}%
  \BibitemOpen
  \bibfield  {author} {\bibinfo {author} {\bibfnamefont {R.}~\bibnamefont
  {Dreyfus}}, \bibinfo {author} {\bibfnamefont {M.~E.}\ \bibnamefont
  {Leunissen}}, \bibinfo {author} {\bibfnamefont {R.}~\bibnamefont {Sha}},
  \bibinfo {author} {\bibfnamefont {A.}~\bibnamefont {Tkachenko}}, \bibinfo
  {author} {\bibfnamefont {N.~C.}\ \bibnamefont {Seeman}}, \bibinfo {author}
  {\bibfnamefont {D.~J.}\ \bibnamefont {Pine}}, \ and\ \bibinfo {author}
  {\bibfnamefont {P.~M.}\ \bibnamefont {Chaikin}},\ }\href {\doibase
  10.1103/PhysRevE.81.041404} {\bibfield  {journal} {\bibinfo  {journal} {Phys.
  Rev. E}\ }\textbf {\bibinfo {volume} {81}},\ \bibinfo {pages} {041404}
  (\bibinfo {year} {2010})}\BibitemShut {NoStop}%
\bibitem [{\citenamefont {Xu}\ \emph {et~al.}(2011)\citenamefont {Xu},
  \citenamefont {Feng}, \citenamefont {Sha}, \citenamefont {Seeman},\ and\
  \citenamefont {Chaikin}}]{chaikin-subdiffusion}%
  \BibitemOpen
  \bibfield  {author} {\bibinfo {author} {\bibfnamefont {Q.}~\bibnamefont
  {Xu}}, \bibinfo {author} {\bibfnamefont {L.}~\bibnamefont {Feng}}, \bibinfo
  {author} {\bibfnamefont {R.}~\bibnamefont {Sha}}, \bibinfo {author}
  {\bibfnamefont {N.~C.}\ \bibnamefont {Seeman}}, \ and\ \bibinfo {author}
  {\bibfnamefont {P.~M.}\ \bibnamefont {Chaikin}},\ }\href {\doibase
  10.1103/PhysRevLett.106.228102} {\bibfield  {journal} {\bibinfo  {journal}
  {Phys. Rev. Lett.}\ }\textbf {\bibinfo {volume} {106}},\ \bibinfo {pages}
  {228102} (\bibinfo {year} {2011})}\BibitemShut {NoStop}%
\bibitem [{\citenamefont {Rogers}\ \emph {et~al.}(2013)\citenamefont {Rogers},
  \citenamefont {Sinno},\ and\ \citenamefont {Crocker}}]{rogers2013kinetics}%
  \BibitemOpen
  \bibfield  {author} {\bibinfo {author} {\bibfnamefont {W.~B.}\ \bibnamefont
  {Rogers}}, \bibinfo {author} {\bibfnamefont {T.}~\bibnamefont {Sinno}}, \
  and\ \bibinfo {author} {\bibfnamefont {J.~C.}\ \bibnamefont {Crocker}},\
  }\href@noop {} {\bibfield  {journal} {\bibinfo  {journal} {Soft Matter}\
  }\textbf {\bibinfo {volume} {9}},\ \bibinfo {pages} {6412} (\bibinfo {year}
  {2013})}\BibitemShut {NoStop}%
\bibitem [{\citenamefont {Kim}\ \emph {et~al.}(2006)\citenamefont {Kim},
  \citenamefont {Biancaniello},\ and\ \citenamefont
  {Crocker}}]{KimCrockerLang2006}%
  \BibitemOpen
  \bibfield  {author} {\bibinfo {author} {\bibfnamefont {A.~J.}\ \bibnamefont
  {Kim}}, \bibinfo {author} {\bibfnamefont {P.~L.}\ \bibnamefont
  {Biancaniello}}, \ and\ \bibinfo {author} {\bibfnamefont {J.~C.}\
  \bibnamefont {Crocker}},\ }\href {\doibase 10.1021/la0528955} {\bibfield
  {journal} {\bibinfo  {journal} {Langmuir}\ }\textbf {\bibinfo {volume}
  {22}},\ \bibinfo {pages} {1991} (\bibinfo {year} {2006})},\ \Eprint
  {http://arxiv.org/abs/http://pubs.acs.org/doi/pdf/10.1021/la0528955}
  {http://pubs.acs.org/doi/pdf/10.1021/la0528955} \BibitemShut {NoStop}%
\bibitem [{\citenamefont {Wang}\ \emph
  {et~al.}(2015{\natexlab{a}})\citenamefont {Wang}, \citenamefont {Wang},
  \citenamefont {Zheng}, \citenamefont {Ducrot}, \citenamefont {Yodh},
  \citenamefont {Weck},\ and\ \citenamefont {Pine}}]{wang2015crystallization}%
  \BibitemOpen
  \bibfield  {author} {\bibinfo {author} {\bibfnamefont {Y.}~\bibnamefont
  {Wang}}, \bibinfo {author} {\bibfnamefont {Y.}~\bibnamefont {Wang}}, \bibinfo
  {author} {\bibfnamefont {X.}~\bibnamefont {Zheng}}, \bibinfo {author}
  {\bibfnamefont {{\'E}.}~\bibnamefont {Ducrot}}, \bibinfo {author}
  {\bibfnamefont {J.~S.}\ \bibnamefont {Yodh}}, \bibinfo {author}
  {\bibfnamefont {M.}~\bibnamefont {Weck}}, \ and\ \bibinfo {author}
  {\bibfnamefont {D.~J.}\ \bibnamefont {Pine}},\ }\href@noop {} {\bibfield
  {journal} {\bibinfo  {journal} {Nat. Commun.}\ }\textbf {\bibinfo {volume}
  {6}},\ \bibinfo {pages} {7253} (\bibinfo {year}
  {2015}{\natexlab{a}})}\BibitemShut {NoStop}%
\bibitem [{\citenamefont {Wang}\ \emph
  {et~al.}(2015{\natexlab{b}})\citenamefont {Wang}, \citenamefont {Wang},
  \citenamefont {Zheng}, \citenamefont {Ducrot}, \citenamefont {Lee},
  \citenamefont {Yi}, \citenamefont {Weck},\ and\ \citenamefont
  {Pine}}]{wang2015synthetic}%
  \BibitemOpen
  \bibfield  {author} {\bibinfo {author} {\bibfnamefont {Y.}~\bibnamefont
  {Wang}}, \bibinfo {author} {\bibfnamefont {Y.}~\bibnamefont {Wang}}, \bibinfo
  {author} {\bibfnamefont {X.}~\bibnamefont {Zheng}}, \bibinfo {author}
  {\bibfnamefont {{\'E}.}~\bibnamefont {Ducrot}}, \bibinfo {author}
  {\bibfnamefont {M.~G.}\ \bibnamefont {Lee}}, \bibinfo {author} {\bibfnamefont
  {G.-R.}\ \bibnamefont {Yi}}, \bibinfo {author} {\bibfnamefont
  {M.}~\bibnamefont {Weck}}, \ and\ \bibinfo {author} {\bibfnamefont {D.~J.}\
  \bibnamefont {Pine}},\ }\href@noop {} {\bibfield  {journal} {\bibinfo
  {journal} {J. Am. Chem. Soc.}\ }\textbf {\bibinfo {volume} {137}},\ \bibinfo
  {pages} {10760} (\bibinfo {year} {2015}{\natexlab{b}})}\BibitemShut {NoStop}%
\bibitem [{\citenamefont {Girard}\ \emph {et~al.}(2019)\citenamefont {Girard},
  \citenamefont {Wang}, \citenamefont {Du}, \citenamefont {Das}, \citenamefont
  {Huang}, \citenamefont {Dravid}, \citenamefont {Lee}, \citenamefont
  {Mirkin},\ and\ \citenamefont {Olvera de~la Cruz}}]{Girard1174}%
  \BibitemOpen
  \bibfield  {author} {\bibinfo {author} {\bibfnamefont {M.}~\bibnamefont
  {Girard}}, \bibinfo {author} {\bibfnamefont {S.}~\bibnamefont {Wang}},
  \bibinfo {author} {\bibfnamefont {J.~S.}\ \bibnamefont {Du}}, \bibinfo
  {author} {\bibfnamefont {A.}~\bibnamefont {Das}}, \bibinfo {author}
  {\bibfnamefont {Z.}~\bibnamefont {Huang}}, \bibinfo {author} {\bibfnamefont
  {V.~P.}\ \bibnamefont {Dravid}}, \bibinfo {author} {\bibfnamefont
  {B.}~\bibnamefont {Lee}}, \bibinfo {author} {\bibfnamefont {C.~A.}\
  \bibnamefont {Mirkin}}, \ and\ \bibinfo {author} {\bibfnamefont
  {M.}~\bibnamefont {Olvera de~la Cruz}},\ }\href {\doibase
  10.1126/science.aaw8237} {\bibfield  {journal} {\bibinfo  {journal}
  {Science}\ }\textbf {\bibinfo {volume} {364}},\ \bibinfo {pages} {1174}
  (\bibinfo {year} {2019})},\ \Eprint
  {http://arxiv.org/abs/https://science.sciencemag.org/content/364/6446/1174.full.pdf}
  {https://science.sciencemag.org/content/364/6446/1174.full.pdf} \BibitemShut
  {NoStop}%
\bibitem [{\citenamefont {Lee-Thorp}\ and\ \citenamefont
  {Holmes-Cerfon}(2018)}]{C8SM01430B}%
  \BibitemOpen
  \bibfield  {author} {\bibinfo {author} {\bibfnamefont {J.~P.}\ \bibnamefont
  {Lee-Thorp}}\ and\ \bibinfo {author} {\bibfnamefont {M.}~\bibnamefont
  {Holmes-Cerfon}},\ }\href {\doibase 10.1039/C8SM01430B} {\bibfield  {journal}
  {\bibinfo  {journal} {Soft Matter}\ }\textbf {\bibinfo {volume} {14}},\
  \bibinfo {pages} {8147} (\bibinfo {year} {2018})}\BibitemShut {NoStop}%
\bibitem [{\citenamefont {Delguste}\ \emph {et~al.}(2018)\citenamefont
  {Delguste}, \citenamefont {Zeippen}, \citenamefont {Machiels}, \citenamefont
  {Mast}, \citenamefont {Gillet},\ and\ \citenamefont
  {Alsteens}}]{Delgusteeaat1273}%
  \BibitemOpen
  \bibfield  {author} {\bibinfo {author} {\bibfnamefont {M.}~\bibnamefont
  {Delguste}}, \bibinfo {author} {\bibfnamefont {C.}~\bibnamefont {Zeippen}},
  \bibinfo {author} {\bibfnamefont {B.}~\bibnamefont {Machiels}}, \bibinfo
  {author} {\bibfnamefont {J.}~\bibnamefont {Mast}}, \bibinfo {author}
  {\bibfnamefont {L.}~\bibnamefont {Gillet}}, \ and\ \bibinfo {author}
  {\bibfnamefont {D.}~\bibnamefont {Alsteens}},\ }\href {\doibase
  10.1126/sciadv.aat1273} {\bibfield  {journal} {\bibinfo  {journal} {Sci.
  Adv.}\ }\textbf {\bibinfo {volume} {4}},\ \bibinfo {pages} {eaat1273}
  (\bibinfo {year} {2018})},\ \Eprint
  {http://arxiv.org/abs/http://advances.sciencemag.org/content/4/8/eaat1273.full.pdf}
  {http://advances.sciencemag.org/content/4/8/eaat1273.full.pdf} \BibitemShut
  {NoStop}%
\bibitem [{\citenamefont {Parveen}\ \emph {et~al.}(2017)\citenamefont
  {Parveen}, \citenamefont {Block}, \citenamefont {Zhdanov}, \citenamefont
  {Rydell},\ and\ \citenamefont {Hook}}]{parveendetachment}%
  \BibitemOpen
  \bibfield  {author} {\bibinfo {author} {\bibfnamefont {N.}~\bibnamefont
  {Parveen}}, \bibinfo {author} {\bibfnamefont {S.}~\bibnamefont {Block}},
  \bibinfo {author} {\bibfnamefont {V.~P.}\ \bibnamefont {Zhdanov}}, \bibinfo
  {author} {\bibfnamefont {G.~E.}\ \bibnamefont {Rydell}}, \ and\ \bibinfo
  {author} {\bibfnamefont {F.}~\bibnamefont {Hook}},\ }\href@noop {} {\bibfield
   {journal} {\bibinfo  {journal} {Langmuir}\ }\textbf {\bibinfo {volume}
  {33}},\ \bibinfo {pages} {4049} (\bibinfo {year} {2017})}\BibitemShut
  {NoStop}%
\bibitem [{\citenamefont {Vahey}\ and\ \citenamefont
  {Fletcher}(2019)}]{vahey2019influenza}%
  \BibitemOpen
  \bibfield  {author} {\bibinfo {author} {\bibfnamefont {M.~D.}\ \bibnamefont
  {Vahey}}\ and\ \bibinfo {author} {\bibfnamefont {D.~A.}\ \bibnamefont
  {Fletcher}},\ }\href@noop {} {\bibfield  {journal} {\bibinfo  {journal}
  {eLife}\ }\textbf {\bibinfo {volume} {8}},\ \bibinfo {pages} {e43764}
  (\bibinfo {year} {2019})}\BibitemShut {NoStop}%
\bibitem [{\citenamefont {Kariuki}\ \emph {et~al.}(2019)\citenamefont
  {Kariuki}, \citenamefont {Marin-Menendez}, \citenamefont {Introini},
  \citenamefont {Ravenhill}, \citenamefont {Lin}, \citenamefont {Macharia},
  \citenamefont {Makale}, \citenamefont {Tendwa}, \citenamefont {Nyamu},
  \citenamefont {Kotar}, \citenamefont {Carrasquilla}, \citenamefont {Rowe},
  \citenamefont {Rockett}, \citenamefont {Kwiatkowski}, \citenamefont {Weekes},
  \citenamefont {Cicuta}, \citenamefont {Williams},\ and\ \citenamefont
  {Rayner}}]{cicuta2019}%
  \BibitemOpen
  \bibfield  {author} {\bibinfo {author} {\bibfnamefont {S.}~\bibnamefont
  {Kariuki}}, \bibinfo {author} {\bibfnamefont {A.}~\bibnamefont
  {Marin-Menendez}}, \bibinfo {author} {\bibfnamefont {V.}~\bibnamefont
  {Introini}}, \bibinfo {author} {\bibfnamefont {B.}~\bibnamefont {Ravenhill}},
  \bibinfo {author} {\bibfnamefont {Y.-C.}\ \bibnamefont {Lin}}, \bibinfo
  {author} {\bibfnamefont {A.}~\bibnamefont {Macharia}}, \bibinfo {author}
  {\bibfnamefont {J.}~\bibnamefont {Makale}}, \bibinfo {author} {\bibfnamefont
  {M.}~\bibnamefont {Tendwa}}, \bibinfo {author} {\bibfnamefont
  {W.}~\bibnamefont {Nyamu}}, \bibinfo {author} {\bibfnamefont
  {J.}~\bibnamefont {Kotar}}, \bibinfo {author} {\bibfnamefont
  {M.}~\bibnamefont {Carrasquilla}}, \bibinfo {author} {\bibfnamefont {J.~A.}\
  \bibnamefont {Rowe}}, \bibinfo {author} {\bibfnamefont {K.}~\bibnamefont
  {Rockett}}, \bibinfo {author} {\bibfnamefont {D.}~\bibnamefont
  {Kwiatkowski}}, \bibinfo {author} {\bibfnamefont {M.}~\bibnamefont {Weekes}},
  \bibinfo {author} {\bibfnamefont {P.}~\bibnamefont {Cicuta}}, \bibinfo
  {author} {\bibfnamefont {T.}~\bibnamefont {Williams}}, \ and\ \bibinfo
  {author} {\bibfnamefont {J.}~\bibnamefont {Rayner}},\ }\href {\doibase
  10.1101/475442} {\bibfield  {journal} {\bibinfo  {journal} {bioRxiv}\ }
  (\bibinfo {year} {2019}),\ 10.1101/475442}\BibitemShut {NoStop}%
\bibitem [{\citenamefont {Bachmann}\ \emph {et~al.}(2016)\citenamefont
  {Bachmann}, \citenamefont {Petitzon},\ and\ \citenamefont
  {Mognetti}}]{PetitzonSoftMatter2016}%
  \BibitemOpen
  \bibfield  {author} {\bibinfo {author} {\bibfnamefont {S.~J.}\ \bibnamefont
  {Bachmann}}, \bibinfo {author} {\bibfnamefont {M.}~\bibnamefont {Petitzon}},
  \ and\ \bibinfo {author} {\bibfnamefont {B.~M.}\ \bibnamefont {Mognetti}},\
  }\href@noop {} {\bibfield  {journal} {\bibinfo  {journal} {Soft matter}\
  }\textbf {\bibinfo {volume} {12}},\ \bibinfo {pages} {9585} (\bibinfo {year}
  {2016})}\BibitemShut {NoStop}%
\bibitem [{\citenamefont {Lanfranco}\ \emph {et~al.}(2019)\citenamefont
  {Lanfranco}, \citenamefont {Jana}, \citenamefont {Tunesi}, \citenamefont
  {Cicuta}, \citenamefont {Mognetti}, \citenamefont {Di~Michele},\ and\
  \citenamefont {Bruylants}}]{lanfranco2019kinetics}%
  \BibitemOpen
  \bibfield  {author} {\bibinfo {author} {\bibfnamefont {R.}~\bibnamefont
  {Lanfranco}}, \bibinfo {author} {\bibfnamefont {P.~K.}\ \bibnamefont {Jana}},
  \bibinfo {author} {\bibfnamefont {L.}~\bibnamefont {Tunesi}}, \bibinfo
  {author} {\bibfnamefont {P.}~\bibnamefont {Cicuta}}, \bibinfo {author}
  {\bibfnamefont {B.~M.}\ \bibnamefont {Mognetti}}, \bibinfo {author}
  {\bibfnamefont {L.}~\bibnamefont {Di~Michele}}, \ and\ \bibinfo {author}
  {\bibfnamefont {G.}~\bibnamefont {Bruylants}},\ }\href@noop {} {\bibfield
  {journal} {\bibinfo  {journal} {Langmuir}\ }\textbf {\bibinfo {volume}
  {35}},\ \bibinfo {pages} {2002} (\bibinfo {year} {2019})}\BibitemShut
  {NoStop}%
\bibitem [{\citenamefont {Parolini}\ \emph {et~al.}(2016)\citenamefont
  {Parolini}, \citenamefont {Kotar}, \citenamefont {Di~Michele},\ and\
  \citenamefont {Mognetti}}]{ParoliniACSNano2016}%
  \BibitemOpen
  \bibfield  {author} {\bibinfo {author} {\bibfnamefont {L.}~\bibnamefont
  {Parolini}}, \bibinfo {author} {\bibfnamefont {J.}~\bibnamefont {Kotar}},
  \bibinfo {author} {\bibfnamefont {L.}~\bibnamefont {Di~Michele}}, \ and\
  \bibinfo {author} {\bibfnamefont {B.~M.}\ \bibnamefont {Mognetti}},\
  }\href@noop {} {\bibfield  {journal} {\bibinfo  {journal} {ACS Nano}\
  }\textbf {\bibinfo {volume} {10}},\ \bibinfo {pages} {2392} (\bibinfo {year}
  {2016})}\BibitemShut {NoStop}%
\bibitem [{\citenamefont {Jana}\ and\ \citenamefont
  {Mognetti}(2019)}]{jana2019surface}%
  \BibitemOpen
  \bibfield  {author} {\bibinfo {author} {\bibfnamefont {P.~K.}\ \bibnamefont
  {Jana}}\ and\ \bibinfo {author} {\bibfnamefont {B.~M.}\ \bibnamefont
  {Mognetti}},\ }\href@noop {} {\bibfield  {journal} {\bibinfo  {journal}
  {Nanoscale}\ }\textbf {\bibinfo {volume} {11}},\ \bibinfo {pages} {5450}
  (\bibinfo {year} {2019})}\BibitemShut {NoStop}%
\bibitem [{\citenamefont {Ho}\ \emph {et~al.}(2009)\citenamefont {Ho},
  \citenamefont {Zimmermann}, \citenamefont {Dehmelt}, \citenamefont
  {Steinbach}, \citenamefont {Erdmann}, \citenamefont {Severin}, \citenamefont
  {Falter},\ and\ \citenamefont {Gaub}}]{Ho_BiophJ_2009}%
  \BibitemOpen
  \bibfield  {author} {\bibinfo {author} {\bibfnamefont {D.}~\bibnamefont
  {Ho}}, \bibinfo {author} {\bibfnamefont {J.~L.}\ \bibnamefont {Zimmermann}},
  \bibinfo {author} {\bibfnamefont {F.~A.}\ \bibnamefont {Dehmelt}}, \bibinfo
  {author} {\bibfnamefont {U.}~\bibnamefont {Steinbach}}, \bibinfo {author}
  {\bibfnamefont {M.}~\bibnamefont {Erdmann}}, \bibinfo {author} {\bibfnamefont
  {P.}~\bibnamefont {Severin}}, \bibinfo {author} {\bibfnamefont
  {K.}~\bibnamefont {Falter}}, \ and\ \bibinfo {author} {\bibfnamefont {H.~E.}\
  \bibnamefont {Gaub}},\ }\href@noop {} {\bibfield  {journal} {\bibinfo
  {journal} {Biophys.\ J.}\ }\textbf {\bibinfo {volume} {97}},\ \bibinfo
  {pages} {3158} (\bibinfo {year} {2009})}\BibitemShut {NoStop}%
\bibitem [{\citenamefont {Chang}\ and\ \citenamefont
  {Hammer}(1996)}]{chang1996influence}%
  \BibitemOpen
  \bibfield  {author} {\bibinfo {author} {\bibfnamefont {K.-C.}\ \bibnamefont
  {Chang}}\ and\ \bibinfo {author} {\bibfnamefont {D.~A.}\ \bibnamefont
  {Hammer}},\ }\href@noop {} {\bibfield  {journal} {\bibinfo  {journal}
  {Langmuir}\ }\textbf {\bibinfo {volume} {12}},\ \bibinfo {pages} {2271}
  (\bibinfo {year} {1996})}\BibitemShut {NoStop}%
\bibitem [{\citenamefont {Chang}\ \emph {et~al.}(2000)\citenamefont {Chang},
  \citenamefont {Tees},\ and\ \citenamefont {Hammer}}]{chang2000state}%
  \BibitemOpen
  \bibfield  {author} {\bibinfo {author} {\bibfnamefont {K.-C.}\ \bibnamefont
  {Chang}}, \bibinfo {author} {\bibfnamefont {D.~F.~J.}\ \bibnamefont {Tees}},
  \ and\ \bibinfo {author} {\bibfnamefont {D.~A.}\ \bibnamefont {Hammer}},\
  }\href@noop {} {\bibfield  {journal} {\bibinfo  {journal} {Proc. Natl. Acad.
  Sci. USA}\ }\textbf {\bibinfo {volume} {97}},\ \bibinfo {pages} {11262}
  (\bibinfo {year} {2000})}\BibitemShut {NoStop}%
\bibitem [{\citenamefont {Shah}\ \emph {et~al.}(2011)\citenamefont {Shah},
  \citenamefont {Liu}, \citenamefont {Hu},\ and\ \citenamefont
  {Gao}}]{shah2011modeling}%
  \BibitemOpen
  \bibfield  {author} {\bibinfo {author} {\bibfnamefont {S.}~\bibnamefont
  {Shah}}, \bibinfo {author} {\bibfnamefont {Y.}~\bibnamefont {Liu}}, \bibinfo
  {author} {\bibfnamefont {W.}~\bibnamefont {Hu}}, \ and\ \bibinfo {author}
  {\bibfnamefont {J.}~\bibnamefont {Gao}},\ }\href@noop {} {\bibfield
  {journal} {\bibinfo  {journal} {J. Nanosci. Nanotechnol.}\ }\textbf {\bibinfo
  {volume} {11}},\ \bibinfo {pages} {919} (\bibinfo {year} {2011})}\BibitemShut
  {NoStop}%
\bibitem [{\citenamefont {SantaLucia}(1998)}]{santalucia}%
  \BibitemOpen
  \bibfield  {author} {\bibinfo {author} {\bibfnamefont {J.}~\bibnamefont
  {SantaLucia}},\ }\href {http://www.pnas.org/content/95/4/1460.abstract}
  {\bibfield  {journal} {\bibinfo  {journal} {Proc. Natl. Acad. Sci. USA}\
  }\textbf {\bibinfo {volume} {95}},\ \bibinfo {pages} {1460} (\bibinfo {year}
  {1998})},\ \Eprint
  {http://arxiv.org/abs/http://www.pnas.org/content/95/4/1460.full.pdf+html}
  {http://www.pnas.org/content/95/4/1460.full.pdf+html} \BibitemShut {NoStop}%
\bibitem [{\citenamefont {Zadeh}\ \emph {et~al.}(2011)\citenamefont {Zadeh},
  \citenamefont {Steenberg}, \citenamefont {Bois}, \citenamefont {Wolfe},
  \citenamefont {Pierce}, \citenamefont {Khan}, \citenamefont {Dirks},\ and\
  \citenamefont {Pierce}}]{zadeh2011nupack}%
  \BibitemOpen
  \bibfield  {author} {\bibinfo {author} {\bibfnamefont {J.~N.}\ \bibnamefont
  {Zadeh}}, \bibinfo {author} {\bibfnamefont {C.~D.}\ \bibnamefont
  {Steenberg}}, \bibinfo {author} {\bibfnamefont {J.~S.}\ \bibnamefont {Bois}},
  \bibinfo {author} {\bibfnamefont {B.~R.}\ \bibnamefont {Wolfe}}, \bibinfo
  {author} {\bibfnamefont {M.~B.}\ \bibnamefont {Pierce}}, \bibinfo {author}
  {\bibfnamefont {A.~R.}\ \bibnamefont {Khan}}, \bibinfo {author}
  {\bibfnamefont {R.~M.}\ \bibnamefont {Dirks}}, \ and\ \bibinfo {author}
  {\bibfnamefont {N.~A.}\ \bibnamefont {Pierce}},\ }\href@noop {} {\bibfield
  {journal} {\bibinfo  {journal} {J. Comp. Chem.}\ }\textbf {\bibinfo {volume}
  {32}},\ \bibinfo {pages} {170} (\bibinfo {year} {2011})}\BibitemShut
  {NoStop}%
\bibitem [{\citenamefont {Mognetti}\ \emph {et~al.}(2012)\citenamefont
  {Mognetti}, \citenamefont {Leunissen},\ and\ \citenamefont
  {Frenkel}}]{MognettiSoftMatt2012}%
  \BibitemOpen
  \bibfield  {author} {\bibinfo {author} {\bibfnamefont {B.~M.}\ \bibnamefont
  {Mognetti}}, \bibinfo {author} {\bibfnamefont {M.~E.}\ \bibnamefont
  {Leunissen}}, \ and\ \bibinfo {author} {\bibfnamefont {D.}~\bibnamefont
  {Frenkel}},\ }\href {\doibase 10.1039/C2SM06635A} {\bibfield  {journal}
  {\bibinfo  {journal} {Soft Matter}\ }\textbf {\bibinfo {volume} {8}},\
  \bibinfo {pages} {2213} (\bibinfo {year} {2012})}\BibitemShut {NoStop}%
\bibitem [{\citenamefont {Varilly}\ \emph {et~al.}(2012)\citenamefont
  {Varilly}, \citenamefont {Angioletti-Uberti}, \citenamefont {Mognetti},\ and\
  \citenamefont {Frenkel}}]{VarillyJCP2012}%
  \BibitemOpen
  \bibfield  {author} {\bibinfo {author} {\bibfnamefont {P.}~\bibnamefont
  {Varilly}}, \bibinfo {author} {\bibfnamefont {S.}~\bibnamefont
  {Angioletti-Uberti}}, \bibinfo {author} {\bibfnamefont {B.~M.}\ \bibnamefont
  {Mognetti}}, \ and\ \bibinfo {author} {\bibfnamefont {D.}~\bibnamefont
  {Frenkel}},\ }\href {\doibase http://arxiv.org/abs/1205.6921} {\bibfield
  {journal} {\bibinfo  {journal} {J. Chem. Phys.}\ }\textbf {\bibinfo {volume}
  {137}},\ \bibinfo {pages} {094108} (\bibinfo {year} {2012})}\BibitemShut
  {NoStop}%
\bibitem [{\citenamefont {Peck}\ \emph {et~al.}(2015)\citenamefont {Peck},
  \citenamefont {Liu}, \citenamefont {Spence}, \citenamefont {Shaw},
  \citenamefont {Davis}, \citenamefont {Destecroix},\ and\ \citenamefont
  {Smith}}]{peck2015rapid}%
  \BibitemOpen
  \bibfield  {author} {\bibinfo {author} {\bibfnamefont {E.~M.}\ \bibnamefont
  {Peck}}, \bibinfo {author} {\bibfnamefont {W.}~\bibnamefont {Liu}}, \bibinfo
  {author} {\bibfnamefont {G.~T.}\ \bibnamefont {Spence}}, \bibinfo {author}
  {\bibfnamefont {S.~K.}\ \bibnamefont {Shaw}}, \bibinfo {author}
  {\bibfnamefont {A.~P.}\ \bibnamefont {Davis}}, \bibinfo {author}
  {\bibfnamefont {H.}~\bibnamefont {Destecroix}}, \ and\ \bibinfo {author}
  {\bibfnamefont {B.~D.}\ \bibnamefont {Smith}},\ }\href@noop {} {\bibfield
  {journal} {\bibinfo  {journal} {J. Am. Chem. Soc.}\ }\textbf {\bibinfo
  {volume} {137}},\ \bibinfo {pages} {8668} (\bibinfo {year}
  {2015})}\BibitemShut {NoStop}%
\bibitem [{\citenamefont {Randeria}\ \emph {et~al.}(2015)\citenamefont
  {Randeria}, \citenamefont {Jones}, \citenamefont {Kohlstedt}, \citenamefont
  {Banga}, \citenamefont {Olvera de~la Cruz}, \citenamefont {Schatz},\ and\
  \citenamefont {Mirkin}}]{randeria2015controls}%
  \BibitemOpen
  \bibfield  {author} {\bibinfo {author} {\bibfnamefont {P.~S.}\ \bibnamefont
  {Randeria}}, \bibinfo {author} {\bibfnamefont {M.~R.}\ \bibnamefont {Jones}},
  \bibinfo {author} {\bibfnamefont {K.~L.}\ \bibnamefont {Kohlstedt}}, \bibinfo
  {author} {\bibfnamefont {R.~J.}\ \bibnamefont {Banga}}, \bibinfo {author}
  {\bibfnamefont {M.}~\bibnamefont {Olvera de~la Cruz}}, \bibinfo {author}
  {\bibfnamefont {G.~C.}\ \bibnamefont {Schatz}}, \ and\ \bibinfo {author}
  {\bibfnamefont {C.~A.}\ \bibnamefont {Mirkin}},\ }\href@noop {} {\bibfield
  {journal} {\bibinfo  {journal} {J. Am. Chem. Soc.}\ }\textbf {\bibinfo
  {volume} {137}},\ \bibinfo {pages} {3486} (\bibinfo {year}
  {2015})}\BibitemShut {NoStop}%
\bibitem [{\citenamefont {Hunter}\ \emph {et~al.}(2011)\citenamefont {Hunter},
  \citenamefont {Edmond}, \citenamefont {Elsesser},\ and\ \citenamefont
  {Weeks}}]{hunter2011tracking}%
  \BibitemOpen
  \bibfield  {author} {\bibinfo {author} {\bibfnamefont {G.~L.}\ \bibnamefont
  {Hunter}}, \bibinfo {author} {\bibfnamefont {K.~V.}\ \bibnamefont {Edmond}},
  \bibinfo {author} {\bibfnamefont {M.~T.}\ \bibnamefont {Elsesser}}, \ and\
  \bibinfo {author} {\bibfnamefont {E.~R.}\ \bibnamefont {Weeks}},\ }\href@noop
  {} {\bibfield  {journal} {\bibinfo  {journal} {Opt. Express}\ }\textbf
  {\bibinfo {volume} {19}},\ \bibinfo {pages} {17189} (\bibinfo {year}
  {2011})}\BibitemShut {NoStop}%
\bibitem [{\citenamefont {Bra\ifmmode~\acute{n}\else \'{n}\fi{}ka}\ and\
  \citenamefont {Heyes}(1994)}]{PhysRevE.50.4810}%
  \BibitemOpen
  \bibfield  {author} {\bibinfo {author} {\bibfnamefont {A.~C.}\ \bibnamefont
  {Bra\ifmmode~\acute{n}\else \'{n}\fi{}ka}}\ and\ \bibinfo {author}
  {\bibfnamefont {D.~M.}\ \bibnamefont {Heyes}},\ }\href {\doibase
  10.1103/PhysRevE.50.4810} {\bibfield  {journal} {\bibinfo  {journal} {Phys.
  Rev. E}\ }\textbf {\bibinfo {volume} {50}},\ \bibinfo {pages} {4810}
  (\bibinfo {year} {1994})}\BibitemShut {NoStop}%
\bibitem [{\citenamefont {Gillespie}(1977)}]{gillespie1977exact}%
  \BibitemOpen
  \bibfield  {author} {\bibinfo {author} {\bibfnamefont {D.~T.}\ \bibnamefont
  {Gillespie}},\ }\href@noop {} {\bibfield  {journal} {\bibinfo  {journal} {The
  journal of physical chemistry}\ }\textbf {\bibinfo {volume} {81}},\ \bibinfo
  {pages} {2340} (\bibinfo {year} {1977})}\BibitemShut {NoStop}%
\bibitem [{\citenamefont {Sch{\"o}neberg}\ and\ \citenamefont
  {No{\'e}}(2013)}]{schoneberg2013readdy}%
  \BibitemOpen
  \bibfield  {author} {\bibinfo {author} {\bibfnamefont {J.}~\bibnamefont
  {Sch{\"o}neberg}}\ and\ \bibinfo {author} {\bibfnamefont {F.}~\bibnamefont
  {No{\'e}}},\ }\href@noop {} {\bibfield  {journal} {\bibinfo  {journal} {PloS
  one}\ }\textbf {\bibinfo {volume} {8}},\ \bibinfo {pages} {e74261} (\bibinfo
  {year} {2013})}\BibitemShut {NoStop}%
\bibitem [{SM()}]{SM}%
  \BibitemOpen
  \bibinfo {note} {See Supplemental Material at [URL will be inserted by
  publisher] for additional details on the model, the simulation algorithm, and
  a figure reporting the number of linkages as a function of the reaction
  rate.}\BibitemShut {Stop}%
\bibitem [{\citenamefont {Jana}\ \emph {et~al.}(2017)\citenamefont {Jana},
  \citenamefont {Alava},\ and\ \citenamefont
  {Zapperi}}]{jana2017irreversibility}%
  \BibitemOpen
  \bibfield  {author} {\bibinfo {author} {\bibfnamefont {P.~K.}\ \bibnamefont
  {Jana}}, \bibinfo {author} {\bibfnamefont {M.~J.}\ \bibnamefont {Alava}}, \
  and\ \bibinfo {author} {\bibfnamefont {S.}~\bibnamefont {Zapperi}},\
  }\href@noop {} {\bibfield  {journal} {\bibinfo  {journal} {Sci. Rep.}\
  }\textbf {\bibinfo {volume} {7}},\ \bibinfo {pages} {45550} (\bibinfo {year}
  {2017})}\BibitemShut {NoStop}%
\bibitem [{\citenamefont {Jana}\ \emph {et~al.}(2018)\citenamefont {Jana},
  \citenamefont {Alava},\ and\ \citenamefont {Zapperi}}]{PhysRevE.98.062607}%
  \BibitemOpen
  \bibfield  {author} {\bibinfo {author} {\bibfnamefont {P.~K.}\ \bibnamefont
  {Jana}}, \bibinfo {author} {\bibfnamefont {M.~J.}\ \bibnamefont {Alava}}, \
  and\ \bibinfo {author} {\bibfnamefont {S.}~\bibnamefont {Zapperi}},\ }\href
  {\doibase 10.1103/PhysRevE.98.062607} {\bibfield  {journal} {\bibinfo
  {journal} {Phys. Rev. E}\ }\textbf {\bibinfo {volume} {98}},\ \bibinfo
  {pages} {062607} (\bibinfo {year} {2018})}\BibitemShut {NoStop}%
\bibitem [{\citenamefont {Zhang}\ and\ \citenamefont
  {Winfree}(2009)}]{zhang2009control}%
  \BibitemOpen
  \bibfield  {author} {\bibinfo {author} {\bibfnamefont {D.~Y.}\ \bibnamefont
  {Zhang}}\ and\ \bibinfo {author} {\bibfnamefont {E.}~\bibnamefont
  {Winfree}},\ }\href@noop {} {\bibfield  {journal} {\bibinfo  {journal} {J.
  Am. Chem. Soc.}\ }\textbf {\bibinfo {volume} {131}},\ \bibinfo {pages}
  {17303} (\bibinfo {year} {2009})}\BibitemShut {NoStop}%
\end{thebibliography}
\end{document}